\documentclass[runningheads,a4paper]{llncs}

\usepackage{verbatim} 
\usepackage{subfig}
\usepackage{url}
\usepackage{graphicx}

\usepackage{color}
\usepackage{listings}
\usepackage{amsmath, amssymb}
\usepackage{algorithm}

\urldef{\mailsa}\path|Witold.Andrzejewski@cs.put.poznan.pl| 
\urldef{\mailsb}\path|a.gramacki@iie.uz.zgora.pl| 
\urldef{\mailsc}\path|j.gramacki@ck.uz.zgora.pl| 
\newcommand{\keywords}[1]{\par\addvspace\baselineskip
\noindent\keywordname\enspace\ignorespaces#1}

\begin{document}

\mainmatter  

\title{Density Estimations for Approximate Query Processing on SIMD Architectures}

\titlerunning{Density Estimations for AQP on SIMD Architectures}

\author{Witold Andrzejewski$^*$, Artur Gramacki$^{**}$ \and Jaros\l{}aw Gramacki$^{***}$}
\authorrunning{Density Estimations for AQP on SIMD Architectures}

\institute{$^*$Pozna\'n{} University of Technology, Institute of Computer Science\\
$^{**}$University of Zielona G\'o{}ra, Institute of Computer Science and Electronics \\
$^{***}$University of Zielona G\'o{}ra, Computer Center
\mailsa\\
\mailsb\\
\mailsc\\
}

\toctitle{Lecture Notes in Computer Science}
\tocauthor{Authors' Instructions}
\maketitle

\begin{abstract}
Approximate query processing (AQP) is an interesting alternative for exact query processing. It is a tool for dealing with the huge data volumes where response time is more important than perfect accuracy (this is typically the case during initial phase of data exploration). There are many techniques for AQP, one of them is based on probability density functions (PDF). PDFs are typically calculated using nonparametric data-driven methods. One of the most popular nonparametric method is the kernel density estimator (KDE). However, a very serious drawback of using KDEs is the large number of calculations required to compute them. The shape of final density function is very sensitive to an entity called bandwidth or smoothing parameter. Calculating it's optimal value is not a trivial task and in general is very time consuming. In this paper we investigate the possibility of utilizing two SIMD architectures: SSE CPU extensions and NVIDIA's CUDA architecture to accelerate finding of the bandwidth. Our experiments show orders of magnitude improvements over a simple sequential implementation of classical algorithms used for that task. 

\keywords{approximate query processing, graphics processing unit, probability density function, nonparametric estimation, kernel estimation, bandwidth selection}

\end{abstract}

\section{Introduction}\label{sec:intro}
The paper is about implementing an \emph{approximate query processing} (AQP) technique with the support of two SIMD architectures: SSE extensions of modern CPUs and Graphics Processing Units (GPUs). We propose modifications of classical algorithms which perform parallel and concurrent computations to accelerate very time-consuming operations while calculating the optimal \emph{kernel density estimators} (KDE), which heavily depends on the so called \emph{bandwidth} or \emph{smoothing parameter}. These estimators estimate probability density functions (PDF) which can be used in the AQP task. 

The perfect situation is if every query that is sent to the database engine could return an exact solution in no more than seconds. However, if the database stores really huge amount of data (as it is the most typical in data warehouses, DW) such a perfect behavior is not a general rule. A lot of theoretical as well as practical works have been done so far in the area of DW optimization \cite{Kozielski:2009}. The research is usually concentrated on such aspects as designing a dedicated logical and physical data structures (novel schemes for indexing, materialized views, partitioning, etc.). Relatively less attention is paid for getting approximate results. Is seems obvious that if we have the following dilemma ``what is better: getting the exact solution in 1 hour or getting only approximate solution in less then seconds'', we probably lean toward the second possibility. Getting only approximate solutions, instead of exact ones, in many practical situations is absolutely acceptable. For example, if our goal is to calculate the total gross income within the whole fiscal year, the roundoff to full thousands is obviously correct approach. There are at least a few different schemes for implementing AQP. We supply a brief review of them in section~\ref{sec:approx-query}. 

In the paper we concentrate on a technique based on analyzing \emph{statistical properties} of data. We use probability density functions which give an elegant and efficient framework for implementing AQP. However, one of the most serious drawback of this approach is that typical kernel nonparametric algorithms used in evaluating optimal PDFs scale quadratically (see section \ref{sec:form-band-sel} for precise mathematical formulas). If data sizes increase, such kernel methods scale poorly. To overcome this problem, two main approaches may be used. In the first one, many authors propose various methods of evaluating \emph{approximate kernel density estimates}. We give a short summary of these methods in section \ref{sec:comp-of-pdf}. In this paper we investigate the second approach where kernel density estimates are evaluating \emph{exactly}. To make this approach practically usable, all the time consuming computations are performed using two modern SIMD architectures. We describe two implementations of each of three different bandwidth finding algorithms. The first implementation (later called \emph{SSE implementation}) utilizes two benefits of modern CPUs: SSE extensions allowing to perform the same instructions on multiple values in parallel as well as multicore architecture allowing to process several threads in parallel on several distinct cores. The second implementation (later called \emph{GPU implementation}) utilizes NVIDIA's CUDA architecture to start multiple threads and perform as many computations concurently or in parallel as possible. The speedups we obtain are in the range of one to two orders of magnitude in comparison with classical sequential CPU-based implementation (later called \emph{Sequential implementation}). 

The remainder of the paper is organized as follows. In section \ref{sec:rel-works} we cover the necessary background material on APQ, computation of PDFs and bandwidth selection. In Section \ref{sec:simd-intro} we supply a brief review of modern SIMD architectures.  In Section \ref{sec:prelim} we turn our attention to give the reader some preliminary informations on PDFs, kernel estimators of PDFs, using PDFs in database area. We also give the very detail mathematical formulas for calculating optimal bandwidth parameters using three different methods. We also propose some modifications of the basic formulas to improve calculations performance. In section \ref{sec:simd} we cover all the necessary details on parallel algorithms for bandwidth selection. In section \ref{sec:algorithms} we show how to utilize the algorithms presented in section \ref{sec:simd}. In section \ref{sec:experiments} we give details on hardware and software used in experiments performed and present the results of the experiments (presented as speedups over different implementations, that is SSE, GPU and sequential ones). In section \ref{sec:c-fw} we conclude our paper. In appendix \ref{sec:jqderiv} we give the details on how we derived some important formulas later used in our algorithms. 

\section{Related works}\label{sec:rel-works}
\subsection{Approximate query processing}\label{sec:approx-query}
Approximate query processing can be done in many different schemes \cite{Garofalakis:2001}. They all assume applying a  kind of preliminary data reduction which gives as a result a synopsis of the original input data. All the following data queries operate on this synopsis instead of the original data. Probably the simplest method of obtaining synopses is sampling. In this case we believe that data after sampling remain still sufficiently representative. This method is often called \emph{numerosity reduction} \cite{Han:2006}. There is also a complementary to numerosity reduction method called \emph{dimensionality reduction}, but this technique is rather not used in the AQP area  \cite{Cunningham:2007,Fodor:2002}. Note also that many commercial RDBMS use sampling in the process of determining the best query execution plans. The synopsis can be built using such techniques as: histograms \cite{Ioannidis:1999}, wavelets \cite{Vitter:1999} or investigating statistical properties of the data \cite{Shanmugasundaram:1999}. The latter approach is investigated in the paper.

\subsection{Computation of probability density functions}\label{sec:comp-of-pdf}
The probability density function (also called \emph{probability distribution}, \emph{density function} or simply \emph{density}) is one of the most important and fundamental concept in statistics and it is widely used in exploratory data analysis. Fast calculation of the optimal PDF is still an interesting scientific problem. There exist a lot of parametric forms of density functions, that is if their shapes can be described by a mathematical formula. A very complete list of probability density functions can be found for example in \cite{Johnson:1994,Johnson:1995}. On the other hand, if the parametric form of the PDF is unknown (or difficult to calculate) one should consider usage of nonparametric methods. 

The task of the PDF estimation is to compute an estimate $\hat{f}$ based on the given $n$ sample points drawn from an unknown population of data with density $f$. One of the most popular nonparametric method is the kernel density estimator, see for example \cite{Wand:1995} for a complete review. 

There are two main computational problems related to KDE. Firstly, the calculation of the estimate $\hat{f}(x,h)$ (or $\hat{f}(x,H)$, see chapter \ref{sec:kde} for shortened explanation on differences beetwen $h$ and $H$) and secondly, estimation of the optimal (in some sense) bandwidth parameter $h$ (or $H$). A plethora of techniques have been proposed for accelerating computational times of the first problem. The naive direct evaluation of the KDE at $m$ evaluation points for $n$ source points requires $O(mn)$ kernel evaluations. Evaluation points can be of course the same as source points and then the computational complexity is $O(n^2)$. The most commonly used method to reduce the computational burden for KDE is to use a technique known as \emph{binning} or \emph{discretising}. In such a case, the density estimation is evaluated at grid points rather than source points. The idea relies on generating grid points (not necessarily equally spaced) of the size $g$, where $g \ll n$ and $g_1 < \cdots < g_m$. Then the original $x_1, \cdots x_n$ source points are replaced by \emph{grid counts} $c_1, \cdots c_g$, where the value of $c_i$ depends on the ''mass of the data'' near $g_i$. Binning strategy reduces the required kernel evaluations to $O(mg)$. Furthermore, if the evaluation points are the same as grid points, the further kernel evaluation from $O(g^2)$ to $O(g)$ is possible (as certain evaluations use the same arguments and don't need to be calculated again). Another approach toward saving computational complexity is based on using Fast Fourier Transformation (FFT), first proposed in \cite{Silverman:1982}. Using FFT requires that the source points are on an evenly spaced grid and then one can evaluate KDE at an evenly spaced grid in $O(n log_n)$. If the source points are irregularly spaced the pre-binning strategy described above should be applied first. The resulting KDE is also evaluated at regular evaluation points. If irregular target points are required, a sort of interpolation based on neighboring evaluation points should be applied. In the FFT-based approach however there is a potential setback connected with an aliasing effect which is not completely bening. This problem is investigated in details in \cite{Hendriks:2003}. Another technique which reduces the computational complexity is based on Fast Gauss Transform (FGT) introduced in \cite{Greengard:1991} and can be viewed as an extension of the Improved Fast Gauss Transform (IFGT) \cite{Yang:2003}. The method is called by the authors $\epsilon-exact$ \cite{Raykar:2006} in the sense that the constant hidden in $O(m+n)$ depends on the desired accuracy which can be chosen arbitrary. 

As for the problem of accelerating computational times for finding the optimal bandwidth $h$ relatively less attention is payed in literature. An attempt at using Message Passing Interface (MPI) was presented in \cite{Lukasik:2007}. In \cite{Silverman:1986} the author gives an FFT-based algorithm for accelerating a method (least-square cross validation one) for finding the optimal bandwidth $h$.

\subsection{Bandwidth selection for kernel density estimates}\label{sec:band-sel}
Bandwidth selection problem is probably the most important one in the KDE area. Fast and correct bandwidth selection is the clue to practical usability of kernel-based density estimates of PDFs. Currently available selectors can be roughly divided into 3 classes \cite{Wand:1995,Silverman:1986,Sheather:2004}. The first class contains very simple and easy to calculate mathematical formulas. They were developed to cover a wide rage of situations, but do not guarantee being close to the optimal bandwidth. They are however willingly and often used as a starting point in more accurate bandwidth selection process. These methods are sometimes called \emph{rules-of-thumb}. The second class contains methods based on \emph{least square} and \emph{cross-validation} ideas and more precise mathematical arguments. But unfortunately they require much more computational effort. However, in reward for it, we get bandwidths more accurate for a wider range of density functions. The method will be abbreviated as \emph{LSCV}. The third class contains methods based on pluging in estimates of some unknown quantities that appear in formulas for the asymptotically optimal bandwidth. The methods are called \emph{plug-in} ones and hereafter will be denoted as \emph{PLUGIN}. These methods are also computationally difficult because there is a need for computation of some functionals and the direct algorithm involves $O(n^2)$ operations. The computational burden can be reduced by using binning strategy as briefly described in section \ref{sec:comp-of-pdf}. 

\section{Single Instruction Multiple Data architectures}\label{sec:simd-intro}

Single Instruction Multiple Data (SIMD) processing systems process multiple streams of data based on a single instruction stream thereby exploiting the data level parallelism. First instroduced as a feature of vector supercomputers such as CDC Star-100 and Thinking Machines CM-1 and CM-2 SIMD processing was later implemented in INTEL's commodity processors. A similar approach, though a little more flexible was implemented in modern GPUs.

\subsection{Parallel data processing on commodity CPUs}\label{sec:parallel}

Starting with Pentium MMX, commodity CPUs have started supporting Single Instruction Multiple Data processing. This was later extended by both Intel\cite{intelarch} and AMD \cite{amdarch1,amdarch2} in subsequent extensions called 3DNow!, SSE, SSE2, SSE3, SSSE3, SSE4, XOP, FMA4, CVT16 (former SSE5) and AVX. CPUs supporting these technologies contain additional registers capable of storing multiple values. These are essentially vectors, which may be processed by specialized instructions as a whole. For example two 4 value vectors stored in two registers may be added by using a single CPU instruction. 

SSE1 (Streaming Simd Extensions) introduced 128bit registers capable of storing four single precision floating point values as well as a set of 70 CPU instructions for processing them. SSE2 added the possibility to store in registers two double precision values instead of four single precision and added additional 144 instructions. SSE3 introduces 13 new instructions, including the ones with capability to perform operations on values stored within the same register. SSE4 added 54 instructions which were (among others) useful for operations performed in HD codecs and for string processing. SSE5 instruction set extension was proposed by AMD on 30 August 2007 as a supplement to the 128-bit SSE core instructions in the AMD64 architecture. In May 2009, AMD replaced SSE5 with three smaller instruction set extensions named as XOP, FMA4, and CVT16, which retain the proposed functionality of SSE5, but encode the instructions differently for better compatibility with Intel's proposed AVX instruction set. AVX extension increases the length of SIMD registers to 256 bits and adds three operand instructions where the destination register is distinct from the source registers.

SIMD is of course not the only level of parallel data processing on modern CPUs. Other solutions used in conujnction with SIMD are: superscalar architecture (multiple execution units and multiple pipelines) as well as multiple cores. Our implementations, aside from SIMD also utilize modern CPUs capability to run multiple threads in parallel on multiple cores. 

In our implementations we primarily use \texttt{\_\_m128} type variables which correspond to 128bit registers holding 4 single precision values.

\subsection{General processing on graphics processing units - GPGPU}\label{sec:gpu}
Rapid development of graphics cards driven by the video game market as well as many professional applications (such as movie special effects) has led to creating devices far more powerful than standard CPUs. Graphics cards are of course more specialized than general-purpose CPUs. Typical processing of graphics algorithms involves performing of the same algorithm steps for many different input values (such as applying geometrical transformations to vertices or computing pixel colors) and is akin to SIMD concept. However, if any algorithm (not necessarily related to computer graphics) may be mapped to such an approach to data processing, it may be efficiently executed on a graphics card. In the beginning, programs utilizing GPUs for general purpose processing used a graphics API such as OpenGL+GLSL or Direct3D+HLSL. This caused some restrictions on the implemented algorithms (lack of the scatter operation) as well as required from the programmer to create some mapping between the algorithm operations and graphics operations. These problems vanished when NVIDIA CUDA and OpenCL were developed. Both of these APIs allow the programmer to completely omit the graphics API and use the standard programming language constructs to create programs for GPUs. In our paper we use the NVIDIA CUDA API, which is closely related to the NVIDIAs graphics cards architecture and therefore allows for some low level optimization.

Let us now roughly describe the NVIDIA GPU architecture and its relation to the CUDA API\cite{cudaprguide}. NVIDIA GPUs are composed of many multiprocessors. Each multiprocessor (SM) is composed of several streaming processors (SP). Each streaming processor is capable of performing logical and integer operations as well as single precision floating point. Groups of SPs on a single SM share a the so-called \emph{warp scheduler} which issues successive instruction or each of the SPs in the group. Consequently, each SP in a group performs the same instruction at the same time (SIMD). Current (2012) graphics cards contain 30 SM with 8 SP and 1 warp scheduler each (NVIDIA GeForce 285GTX), or 16 SM with 32 SP and 2 warp schedulers each (NVIDIA GeForce 580GTX). This shows, that the GPUs are capable of running several hundred threads in parallel (and even more concurrently). Each NVIDIAs graphics card has assigned a compute capability (denoted CC for brevity) version which specifies which features are supported by the given graphics card. Each multiprocessor also contains a small but fast on-chip memory called \emph{the shared memory}. The tasks are not assigned to SMs directly. Instead, the programmer first creates a function called a kernel, which consists of a sequence of operations which need to be performed concurrently in many threads. To distinguish from kernels used in kernel-based density estimates, we will call these functions the \emph{gpu-kernels}. Next, the threads are divided into equally sized \emph{blocks}. A block is a one, two or three dimensional array of at most 1024 threads (or 512 threads on graphics cards with CC$\leq$1.3), where each thread can be uniquely identified by its position in this array. The obtained blocks form the so-called \emph{computation grid}. When the gpu-kernel is started, all of the blocks are automatically distributed among the SMs. Each SM may process more than one block, though one block may be executed at only one SM.  Threads in a single block may communicate by using the same portion of the SMs shared memory. Threads run in different blocks may only communicate through the very slow \emph{global memory} of the graphics card. Synchronization capabilities of the threads are limited. Threads in a block may be synchronized, but global synchronization is not possible, though a (costly) workaround exists. Threads in a block are executed in 32 thread SIMD groups called warps (this is the consequence of having one warp scheduler per several SPs). Consequently all of these threads should perform the same intruction. If the threads with a warp perform different code branches, all branches are serialized, and threads not performing the branch are masked. Perfomance of an implementation is therefore determined by:
 \begin{enumerate}
  \item the number of global memory accesses (the smaller, the better, use shared memory in favor of global memory).
  \item the number of global synchronization points (the smaller, the better, use in-block synchronization in favor of global synchronization).
  \item the parallelism degree (the bigger, the better).
  \item each group of 32 consecutive threads should follow the same code branches.
 \end{enumerate}
There are also several other efficiency guidelines, which do not stem from the above description, but are related to some lower level details of graphics cards hardware:
 \begin{enumerate}
  \item the smaller, the number of conditional code executions the better. 
  \item each thread in a group of 16 (or 32 for graphics cards with CC$\geq$2.0) consecutive threads should follow a conflict-free shared memory access pattern (see~\cite{cudaprguide}).
  \item each thread in a group of 16 (or 32 for graphics cards with CC$\geq$2.0) consecutive threads should use global memory access patterns which allow for coalesced accesses (see~\cite{cudaprguide}).
  \item use single precision floating point computations, if possible.
 \end{enumerate}
All of our implementations comply with the above guidelines. Regarding the last requirement, our implementations use single precision, but in the near future we can expect, that double precision will be efficient on GPUs as well. Our solutions can be then easily ported to support double precision. It should also be noted, that the efficiency of double precision has already increased dramatically between two recent generations of graphics cards (compare for example NVidia GeForce 285GTX and NVidia GeForce 580GTX). 

\section{Mathematical preliminaries}\label{sec:prelim}
In this section we give some preliminary informations on some basic statistical concepts (probability density function and kernel density estimation) as well as how to use them in the database area. Most of this section is devoted to give the precise recipes for calculation of the so called optimal bandwidth which plays the most important role while estimating kernel-based probability density functions. We also propose some slight but important modifications of the reference mathematical formulas for calculating the bandwidth. The modifications play very important role during GPU-based and SSE-based fast algorithm implementations for calculating of the bandwidth. 

\subsection{Probability density function}\label{sec:pdf}
Let $X$ be a random variable and let $f$ be the aforementioned probability density function.  The probability $P$ that a random variable $X$ takes a value in the given interval $[a, b]$ is defined as follows:
\begin{align}\label{eq:pdf-def}
P(a \leq X \leq b) = \int^b_a f(x)dx.
\end{align}

Density function must fulfill two basic properties: must be not negative over the whole domain, that is $f(x) \geq 0$ for every $x$ and the value of the integral (\ref{eq:pdf-def}) must be exactly 1, that is
$P(- \infty \leq X \leq + \infty) = \int^{+\infty}_{-\infty} f(x)dx = 1$. In the context of the main subject of the paper, we also need to know a formula for calculating mean value of the random variable. This is defined as follows:
\begin{align}\label{eq:mean-val}
\mu_X = \int^{+\infty}_{-\infty}xf(x)dx.
\end{align}
The above formulas can be easily generalized into two or more dimensions. In practice, three different cases can be considered: (a) if analytical formula of the PDF is known and the integrals (\ref{eq:pdf-def}) and (\ref{eq:mean-val}) can be solved analytically. In such a case the solution we get is the exact one. (b) if we know the analytical formula of the PDF, but solving the aforementioned integrals is very difficult or even not possible. In such a case one must use methods for numerical integration (a broad family of algorithms exist, for example rectangle rule, trapezoidal rule, Newton-Cotes formula, Gaussian quadrature, Monte Carlo methods and other). (c) if we don't know at all analytical formula of the PDF, then the nonparametric methods for its estimation should be considered. 

\subsection{Kernel density estimation}\label{sec:kde}
Let us assume that we have a set of source points to be a sample from an unknown density function.  \emph{Density estimate} is simply the process of construction of an estimate of the density function (\ref{eq:pdf-def}) from the observed data \cite{Silverman:1986}. There are two main approaches to density estimation: \emph{parametric} and \emph{nonparametric} ones. The first assumes that the data are drawn from a known family of distributions. For example the normal distribution is fully described by only two parameters (mean value $\mu$ and standard deviation $\sigma$). All we need to do is to find estimates of these parameters from the observed data. In the paper this approach is not considered. In the second approach, we assume no preliminary knowledge about a form of a density function. It should be discovered from the set of source points. This approach is sometimes vividly described as \emph{Let the Data Speak for Themselves}. 

A trivial example of a nonparametric PDF is \emph{histogram}. However, its practical usability is rather poor (in the context of using them as PDFs estimators), because in general, and with its basic form, it is difficult to give an optimal number of the bins, their width and the starting point of the first bin. The shape of histogram is very sensitive to these three parameters. In that case, it is much more convenient to use the so called \emph{kernel density estimators} (KDE) which are much more adequate for building nonparametric PDFs. Three classical books on KDEs are \cite{Silverman:1986,Simonoff:1996,Wand:1995}. There are, however, known families of so called optimal histograms (e.g. v-optimal histograms \cite{Jagadish:1998}), but we don't consider them in the paper. We only mention that their construction algorithms are very time and memory consuming. It should be also noted that basic histograms are commonly used in database management systems (DBMSs). This is a feature in cost based optimizers and help the optimizer to decide which SQL execution plan should be used to get the best query execution. 

Now let us consider a random variable $X$ (in general $d$-dimensional) and let assume its probability density function $f$ is not known.  Its estimate, usually denoted by $\hat{f}$, will be determined on the basis of a random sample of size $n$, that is $X_1, X_2, ..., X_n$ (our experimental data). In such a case, the $d$-dimensional kernel density estimator $\hat{f}(x,h)$ of the real density $f(x)$ for random sample $X_1,X_2,\ldots,X_n$ is given by the following formula:
\begin{align}\label{eq:kde-def}
\hat{f}(x,h) = 
n^{-1}\sum_{i=1}^n K_h \left(x-X_i\right)
=
n^{-1}h^{-d}\sum_{i=1}^n K \left( \frac{x-X_i}{h} \right)
\end{align}
where
\begin{align}\label{eq:K_h}
K_h(u) = h^{-1}K(h^{-1}u)
\end{align}
and
\begin{align}\label{eq:gaussian}
K(u) = (2\pi)^{-d/2}exp \left( -\frac{1}{2} u^T u \right)
\end{align}
where $n$ -- number of samples, $d$ -- task dimensionality, $x=(x_1,x_2,\ldots,x_d)^T$, $X_i=(X_{1i},X_{2i},\ldots,X_{di})^T$, $i=1,2,\ldots,n$. $X_{1i}$, $X_{2i}$, $\ldots$, $X_{di}$ denote the consecutive elements of $d$-dimensional vector $X$. $h$ is a positive real number called \emph{smoothing parameter} or \emph{bandwidth}. $K(\cdot)$ is the \emph{kernel function} -- a symetric but not necessarily positive function that integrates to one. In practical applications $K(\cdot)$ is usually the Gaussian normal formula as given in (\ref{eq:gaussian}). Other commonly used kernel functions are Epanechnikov, uniform, triangular, biweight \cite{Li:2007}. One can prove that selection of a particular kernel function is not critical, as all of them guarantee obtaining similar results (Epanechnikov kernel function is theoretically the most effective, but others are only slightly worse, see \cite{Wand:1995} for details). However, the bandwidth is the parameter which exhibits a strong influence on the resulting estimate (shape of the curve). If we have bandwidth $h$ we can determine the estimator  $\hat{f}(x,h)$ of the unknown  $d$-dimensional density function $f(x)$ using the formula (\ref{eq:kde-def}).

Equation (\ref{eq:kde-def}) assumes that the bandwidth $h$ is the scalar quantity, independently of the problem dimensionality. This is the simplest and the least complex variant of the bandwidth. From the other side, the most general and complex version assumes that the so called \emph{bandwidth matrix} $H$ is used instead of the bandwidth scalar $h$. The size of the $H$ matrix is $d \times d$. This matrix is positive definite and symmetric by definition. So, the equivalent of the formula (\ref{eq:kde-def}) is now defined as follows:
\begin{align}\label{eq:kde-def-H}
\hat{f}(x,H) = 
n^{-1}\sum_{i=1}^n K_H \left(x-X_i\right)
=
n^{-1} |H|^{-1/2}\sum_{i=1}^n K \left(H^{-1/2}(x-X_i)\right)
\end{align} 
where
\begin{align}\label{eq:K_H}
K_H(u) = |H|^{-1/2} K(H^{-1/2}u)
\end{align}
and $K(\cdot)$ is defined by (\ref{eq:gaussian}).

Is easy to note that (\ref{eq:K_H}) is not a pure equivalent to (\ref{eq:K_h}) as for univariate case the $1 \times 1$ bandwidth matrix is $H = h^2$. So, now we are dealing with so called 'squared bandwidths'.

A version between the simplest and the most complex is also considered in literature and is based on simplifying the unconstrained bandwidth matrix $H$ to is constrained equivalent where all the off-diagonal entries are zeros by definition. In the paper we do not consider this case. All the possible versions of the bandwidth selectors are investigated in details in \cite{Duong:2004}. Below we only sum up the three main versions of  the bandwidth matrices. 
\begin{align}\label{eq:3_classes_of_H}
h^2I = 
\begin{bmatrix}
h^2 & 0 \\
0 & h^2
\end{bmatrix}, \;\;\;
diagH = 
\begin{bmatrix}
h_1^2 & 0 \\
0 & h_2^2
\end{bmatrix}, \;\;\;
H = 
\begin{bmatrix}
h_1^2 & h_{12} \\
h_{12} & h_2^2
\end{bmatrix}.
\end{align}

Finally, one must remember that nonparametric estimators (to be effective) need more and more data as dimensionality increases. Here, a negative phenomenon called \emph{curse of dimensionality} becomes a real problem. As a consequence, values $\hat{f}(x)$ calculated from (\ref{eq:kde-def}) or (\ref{eq:kde-def-H}) becomes inaccurate. The problem of determining the required sample size needed to achieve a given level of accuracy is studied by some authors. See for example \cite{Silverman:1986,Simonoff:1996}.

\begin{figure}
\centering
\includegraphics[width=1\textwidth]{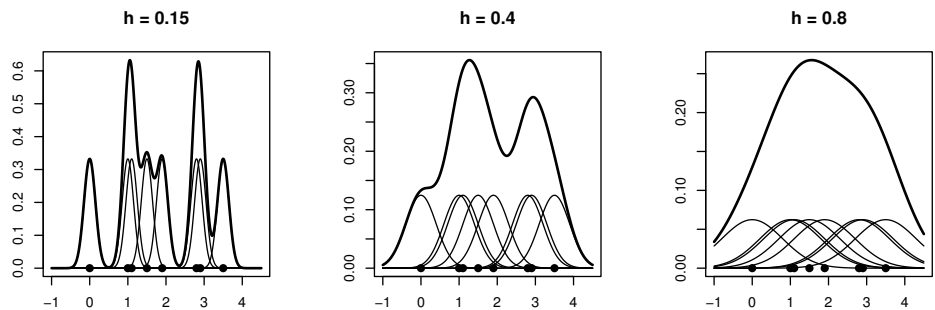}
\caption{An example of using kernel density estimators for determining the probability density function} 
\label{fig-kernel-demo}
\end{figure}

As an example of how KDE works consider a toy dataset of 8 data points $x=\{0, \; 1, \; 1.1, \; 1.5, \; 1.9, \; 2.8, \; 2.9, \;  3.5\}$. Three different PDFs based on these data are depicted in figure \ref{fig-kernel-demo}. It is easy to notice how the bandwidth $h$ influences the shape of the PDF curve. Choosing the best value of $h$ is not a trivial task and this problem was and still is extensively studied in literature \cite{Silverman:1986,Simonoff:1996,Wand:1995}. In Figure \ref{fig-kernel-demo} lines in bold show the estimated PDFs, while normal lines show the shapes of individual kernel functions $K(x)$ (Gaussian normal kernels). Dots represent the data points $x_i$. 

\subsection{Probability density functions in databases}\label{sec:pdf-bd}
A commonly used logical structure in DW area is the multidimensional cube \cite{Han:2006}. The particular dimensions can store both \emph{nominal} (also called \emph{categorical}) as well as \emph{numerical} data (both discrete and continuous). In the case of nominal ones, determining all the possible aggregates is not trivial but feasible task. Of course, in practice usually not all possible aggregates are calculated, as the time and storage requirements would be prohibitive. In contrast, if a dimension stores numerical data one must arbitrary discretize them (by creating separate bins). Calculating an optimal number of bins and their width is not a trivial task. After discretization, the total number of aggregates which are (or will be in the future) possible to create is fixed. To avoid the need for preliminary discretization of data, one can consider using PDFs. As a matter of fact, to calculate them we must read all the data stored (or eventually, only a representative sample of the data). But, what is 
very important, we need to read them only once. Usually we can assume that \emph{statistical properties} of the data remain constant in a period of time (days, months, sometimes even years). So, there is no need to recalculate PDFs after every SQL DML statement executed by a user (INSERT, UPDATE, DELETE). The obvious benefits of PDFs comparing to materialized aggregates is that permanent storing of the former is extremely easy and cheap. Everything what we need to store are either parameters of formula-based PDFs (for example mean value $\mu$ and standard deviation $\sigma$ in the case of Gaussian normal PDF) or output points $\hat{f}(x)$ for nonparametric-based PDFs. 

Now let us consider a relational table where an attribute $column1$ is numerical in nature. If we want to calculate the number of records (\emph{count} aggregate operator is SQL) in the range $a \leq column1 \leq b$ one can use formulas given in section \ref{sec:prelim}. If $n$ is the total number of records, it is trivial to conclude that \emph{approximate} count aggregate can be calculates as
\begin{align}\label{eq:count}
Count = n \cdot \int^b_a f(x)dx. 
\end{align}
Similarly, the sum of all record (\emph{sum} aggregate operator is SQL) in the given range  can be calculates as
\begin{align}\label{eq:sum}
Sum = n \cdot \int^b_a xf(x)dx. 
\end{align}
The similar reasoning leads to conclusion that the product $Sum/Count$ is an equivalent to aggregate SQL's \emph{average} operator. The aforementioned three formulas can be immediately generalised into multidimensional case, that is for 2D case we have
\begin{align}\label{eq:count-sum-2D}
Count = n \cdot \int^d_c \int^b_a f(x,y)dxdy, \nonumber \\
Sum = n \cdot \int^d_c \int^b_a h(x,y) f(x,y)dxdy.
\end{align}
Here $h(x,y)$ is in general a function of two random variables. To calculate for example \emph{sum} of the variable $x$, one have to set $h(x,y)=x$. Above we calculate the sum and the  number of records within logical range $(a \leq column1 \leq b) \; AND \; (c \leq column2 \leq d)$.

The values of the integrals can be done analytically (if the analytical solution exists) or by applying a proper numerical integration method (some of them were mentioned above in section \ref{sec:pdf}). In the case of nonparametric approach, only numerical integration is practicable. In \cite{Gramacki:2010} we give a few practical examples based on real datasets which proof the practical usefulness of using kernel based PDFs for calculating database aggregates.

\subsection{Formulas for bandwidth selection}\label{sec:form-band-sel}
In the following section we present detailed mathematical formulas for calculating bandwidths (optimal in some sense) using the \emph{PLUGIN} and the \emph{LSCV} methods mentioned earlier. The PLUGIN method is designed only for 1D problems, known in literature as \emph{univariate} problems. Contrary to the PLUGIN method, the LSCV method is designed for both 1D as well as nD problems (known in literature as \emph{multivariate}  problems). Three different LSCV versions can be considered, while only two were implemented by the authors. The simplest one (with the smallest computational complexity) assumes that the bandwidth $h$ is the scalar quantity, independently of the problem dimensionality (see equations (\ref{eq:kde-def}) and (\ref{eq:3_classes_of_H})). From the other side, the most computational complex version assumes that the bandwidth matrix $H$ is used, instead of the scalar $h$ (see equation (\ref{eq:kde-def-H})). This matrix is positive definite and symmetric. 

Below, in the next 3 subsections, we give a very condensed recipes for calculating optimal bandwidth $h$ and $H$ using PLUGIN and LSCV approaches described briefly above in section \ref{sec:band-sel}. The LSCV one is presented in two variants: the simplified one for evaluating optimal $h$ (that is if the density estimate is calculated from (\ref{eq:kde-def}); this variant will be called LSCV\_h) and the general multivariate variant of the method (that is if the density estimate is calculated from (\ref{eq:kde-def-H}); this variant will be called LSCV\_H). We comment the individual mathematical formulas very briefly as this is beyond the scope of the paper. Every subsection is prefaced with very short overview of the main ideas of the methods. All the necessary details on the methods as well as details on deriving of particular formulas can be found in many source materials, see for example probably the most often cited books \cite{Wand:1995,Silverman:1986,Simonoff:1996}. 

All the methods presented below determine the optimal bandwidth on the basis of the input random variable and commonly used optimality criterion based on minimization of \emph{mean integrated squared error} (MISE) and its asymptotic approximation (AMISE). 

\subsubsection{PLUGIN}\label{sec:plugin}
This method is used for calculating an optimal bandwidh $h$ for univariate problems, that is applicable to formula (\ref{eq:kde-def}) where $d$ is set to 1. First we calculate the variance and the standard variation estimators of the input data (equations (\ref{eq:PLUGIN-VarianceEstimator}) and (\ref{eq:PLUGIN-StdDevEstimator})). Then we calculate some more complex formulas (equations from (\ref{eq:PLUGIN-Psi8Estimate}) to (\ref{eq:PLUGIN-Psi4Estimate})). The explanations of the essence of them is beyond the scope of the paper and can be found in many books on kernel estimators, for example see the three above-mentioned books. Finally, after completing all the necessary components we can substitute them into equation (\ref{eq:PLUGIN-Final_h}) to get the searched optimal bandwidth $h$.

\begin{enumerate}

 \item Calculate value of variance estimator:
  \begin{align} \label{eq:PLUGIN-VarianceEstimator}
   \hat{V}=\frac{1}{n-1}\sum^{n}_{i=1}X^2_i-\frac{1}{n(n-1)}\left(\sum^n_{i=1}X_i\right)^2.
  \end{align}

 \item Calculate value of standard deviation estimator:
  \begin{align} \label{eq:PLUGIN-StdDevEstimator}
   \hat{\sigma}=\sqrt{\hat{V}}.
  \end{align}

 \item Calculate estimate $\hat{\Psi}^{NS}_8$ of functional $\Psi_8$:
  \begin{align} \label{eq:PLUGIN-Psi8Estimate}
   \hat{\Psi}^{NS}_8=\frac{105}{32\sqrt{\pi}\hat{\sigma}^9}.
  \end{align}

 \item Calculate value of bandwidth of kernel estimator of the function $f^{(4)}$ (4th derivative of the function $f$, that is $f^{(r)}=\frac{d^r}{dx^r}$): 
\begin{align}\label{eq:PLUGIN-g1}
g_1 &= \left( \frac{-2K^6(0)}{\mu_2(K)\hat{\Psi}_8^{NS}n} \right)^{1/9}, \\
K^6(0) &= -\frac{15}{\sqrt{2\pi}},\nonumber \\
\mu_2(K) &= 1.\nonumber
\end{align}

 \item Calculate estimate $\hat{\Psi}_6(g_1)$ of functional $\Psi_6$:
\begin{align}\label{eq:PLUGIN-Psi6Estimate}
\hat{\Psi}_6(g_1) &= 
  \frac{2}{n^2 g_1^7} \sum_{i=1}^{n}\sum_{j=1,i<j}^{n} K^{(6)} 
  \left( \frac{X_i - X_j}{g_1} \right) + nK^{(6)}(0),  \\
K^{6}(x) &= \frac{1}{\sqrt{2\pi}}\left( x^6 - 15x^4 +45x^2 -15 \right) e^{-\frac{1}{2}x^2}. \nonumber
\end{align}

 \item Calculate value of bandwidth of kernel estimator of the function $f^{(2)}$:
\begin{align}\label{eq:PLUGIN-g2}
g_2 &=
\left( \frac{-2K^4(0)}{\mu_2(K)\hat{\Psi}_6(g_1)n} \right)^{1/7},  \\
K^4(0) &= \frac{3}{\sqrt{2\pi}},\nonumber \\
\mu_2(K) &= 1.\nonumber 
\end{align}

\item Calculate estimate $\hat{\Psi}_4(g_2)$ of functional $\Psi_4$:
\begin{align}\label{eq:PLUGIN-Psi4Estimate}
\hat{\Psi}_4(g_2) &= 
  \frac{2}{n^2 g_2^5} \sum_{i=1}^{n}\sum_{j=1,i<j}^{n} K^{(4)} 
  \left( \frac{X_i - X_j}{g_2}\right) + nK^{(4)}(0),\\
K^{4}(x) &= \frac{1}{\sqrt{2\pi}}\left( x^4 - 6x^2 + 3 \right) e^{-\frac{1}{2}x^2}. \nonumber
\end{align}

 \item Calculate the final value of bandwidth $h$:
\begin{align}\label{eq:PLUGIN-Final_h}
h &= \left( \frac{R(K)}{\mu_2(K)^2 \hat{\Psi}_4(g_2) n}\right)^{1/5}, \\
R(K) &= \frac{1}{2\sqrt{\pi}}, \nonumber \\
\mu_2(K) &= 1. \nonumber
\end{align}

\end{enumerate}

\subsubsection{LSCV\_h}\label{sec:lscv_h}
This method is used for calculating an optimal bandwidh $h$ for both univariate and multivariate problems, that is applicable to formula (\ref{eq:kde-def}) where $d$ is set to any integer value qreter or equal to 1. First we calculate the covariance matrix (equation (\ref{eq:LSCV-CovarianceMatrix})), its determinant and its inversion. Then we form an objective function (\ref{eq:LSCV-gh}) which will be minimized according to the searched bandwidth $h$. After determining the search range of the bandwidth $h$ (equation (\ref{eq:LSCV-Zh0})) as well as the starting bandwidth $h$  (equation (\ref{eq:LSCV-h0})), using the ''brute force'' strategy we search for such $h$ which minimizes the objective function (\ref{eq:LSCV-gh}). The performance of the brute force method seems to be acceptable in almost all practical applications. If, however it will turn out to be too slow in practice, a more smart (and faster) method can be used, for example the Golden ratio criterion.
 
\begin{enumerate}
 \item Calculate covariance matrix, given the input data. In columns one can find consecutive $d$-dimensinal vectors of our experimental input data. There are $n$ such vectors:

  \begin{align}\label{eq:LSCV-InputData}
   X=[X_1, X_2, \cdots X_n]=
    \left[\begin{matrix}
     x_{1,1} & x_{1,2} & \cdots & x_{1,n} \\
     x_{2,1} & x_{2,2} & \cdots & x_{2,n} \\
     \vdots  & \vdots  & \ddots & \vdots \\
     x_{d,1} & x_{d,2} & \cdots & x_{d,n} \\
    \end{matrix}\right].
  \end{align}

  Covariance matrix is equal to:

  \begin{align}\label{eq:LSCV-CovarianceMatrix}
   \Sigma=
    \left[\begin{matrix}
     \sigma_{1}^2 & \sigma_{1,2} & \cdots & \sigma_{1,d} \\
     \sigma_{2,1} & \sigma_{2}^2 & \cdots & \sigma_{2,d} \\
     \vdots  & \vdots  & \ddots & \vdots \\
     \sigma_{d,1} & \sigma_{d,2} & \cdots & \sigma_{d}^2 \\
    \end{matrix}\right].
  \end{align}

  where:

   \begin{itemize}
    \item $\sigma^2_i$ - is a variance of each dimension of the random variable,
    \item $\sigma_{i_1,i_2}$ - is a covariance between random variables $i_1$ and $i_2$,
   \end{itemize}

  \begin{align}\label{eq:LSCV-variance}
   \sigma^2_i=\frac{1}{n-1}\sum^n_{j=1}x^2_{i,j}-\frac{1}{n(n-1)}\left(\sum^n_{j=1}x_{i,j}\right)^2,
  \end{align}

  \begin{align}\label{eq:LSCV-covariance}
\sigma_{i_1,i_2}=\frac{1}{n-1}\sum^n_{j=1}x_{i_1,j}x_{i_2,j}-\frac{1}{n(n-1)}\sum^n_{j=1}x_{i_1,j}\sum^n_{j=1}x_{i_2,j}.
  \end{align}

 \item Calculate determinant of the covariance matrix $\Sigma$: $det(\Sigma)$.

 \item Calculate inverse of the covariance matrix $\Sigma$: $\Sigma^{-1}$.

 \item Let $X_i$ be the i-th column of the input matrix $X$. The LSCV\_h method requires to minimize the function $g(h)$:

  \begin{align}\label{eq:LSCV-gh}
   g(h) = h^{-d}\left[2n^{-2}\sum^n_{i=1} \sum^n_{j=1,i<j}T\left( \frac{X_i-X_j}{h}\right) + n^{-1}R(K) \right],
  \end{align}
where
  \begin{align}\label{eq:LSCV-Tx}
   T(u)=(K*K)(u)-2K(u),
  \end{align}

  \begin{align}\label{eq:LSCV-Kx}
   K(u)=(2\pi)^{-d/2} det(\Sigma)^{-1/2} exp\left( -\frac{1}{2}u^T\Sigma^{-1} u\right).
  \end{align}

  \begin{align}\label{eq:LSCV-KKx}
   (K*K)(u)=(4\pi)^{-d/2} det(\Sigma)^{-1/2} exp\left( -\frac{1}{4}u^T\Sigma^{-1} u\right).
  \end{align}

 \item Calculate the approximate value of the bandwidth $h$:

  \begin{align}\label{eq:LSCV-h0}
   h_0 &= \left(\frac{R(K)}{\mu_2(K)^2R(f'')n} \right)^{1/(d+4)} ,\\ \nonumber
   \frac{R(K)}{\mu_2(K)^2} &= \frac{1}{2^d\pi^{d/2}d^2}, \\ \nonumber
   R(f'') &= \frac{d(d+2)}{2^{d+2}\pi^{d/2}}. \nonumber
  \end{align}

 \item Let the range in which we search for minimum of $g(h)$ be heuristically found as:

  \begin{align}\label{eq:LSCV-Zh0}
   Z(h_0)=\left[h_0/4,4h_0 \right].
  \end{align}

  The optimal bandwidth $h$ is equal to:

  \begin{align}\label{eq:LSCV-finalH}
   argmin_{h\in Z(h_0)}g(h).
  \end{align}

\end{enumerate}

\subsubsection{LSCV\_H}\label{sec:lscv_H}
This method is used for calculating the optimal bandwidh matrix $H$ for both univariate and multivariate problems, that is applicable to formula (\ref{eq:kde-def-H}). In this variant of the LSCV method the objective function is defined by equation (\ref{eq:LSCV_MULTI-gh}). Now our goal is to find such the bandwich matrix $H$ which will minimize this objective function. This is the classical nonlinear optimization problem and can be solved by using for example the well known Nelder-Mead method \cite{Nelder:1965}. This method needs a starting matrix which can be calculated from the \emph{rule-of-thumb} heuristic equation (\ref{eq:LSCV-H-start}) taken from \cite{Wand:1995}. 

Two detail notes on evaluating the objective function (\ref{eq:LSCV_MULTI-gh}) are needed here. First, as we said above, Nelder–-Mead method can be used for finding the optimal $H$ bandwidth matrix. This method does not guarantee that in every step the current $H$ is positive--definite. This requirement, however is necessary as the bandwidth $H$ matrix components must be by definition positive scalars. So, the searching for the optimal $H$ must be done over the space of all positive--definite matrices only. Moreover, while calculating the current value of (\ref{eq:LSCV_MULTI-gh}) the inversion of $H$ is needed, hence positive--definite requirement is a must. 

Second, we know that bandwidth matrix $H$ is always a symmetric one (see (\ref{eq:3_classes_of_H})) and its size is $d \times d$. So, only $d(d+1)/2$ independent entries exist. As a consequence there is no need to evaluate the objective function for the full $H$ matrix. It is sufficient to use $vech(H)$, where $vech$ (vector half) operator takes a symmetric $d \times d$ matrix and stacks the lower triangular half into a single vector of length $d(d+1)/2$. That is for an example matrix we have
\begin{align}\label{eq:vech}
A =
\begin{bmatrix}
1 & 4 & 7 \\
2 & 5 & 8 \\
3 & 6 & 9
\end{bmatrix}, \;
vech(A) = 
\begin{bmatrix}
1 & 2 & 3 & 5 & 6 & 9
\end{bmatrix}.
\end{align}

\begin{enumerate}
 \item Let 

\begin{align}\label{eq:LSCV_MULTI-gh}
 g(H)=2 n^{-2}\sum^n_{i=1}\sum^n_{j=1,i<j}T_H(X_i-X_j) + n^{-1} R(K)
\end{align}
where
\begin{align}\label{eq:LSCV_MULTI-Tx}
 T_H(X_i-X_j)=(K*K)_H(X_i-X_j)-2K_H(X_i-X_j), 
\end{align}

\begin{align}\label{eq:LSCV_MULTI-Kx}
 &K_H(X_i-X_j) = \\ \nonumber
 &(2\pi)^{-d/2}|H|^{-1/2}exp\left(-\frac{1}{2}(X_i-X_j)^T H^{-1}(X_i-X_j), \right)
\end{align}

\begin{align}\label{eq:LSCV_MULTI-KKx}
 &(K*K)_H(X_i-X_j) = \\ \nonumber
 &(4\pi)^{-d/2}|H|^{-1/2}exp\left(-\frac{1}{4}(X_i-X_j)^T H^{-1}(X_i-X_j) \right),
\end{align}

\begin{align}\label{eq:LSCV_MULTI-rk}
 R(K)= 2^{-d} \pi^{-d/2} |H|^{-1/2}.
\end{align}

\item Find H which minimizes $g(H)$. Start from ($\Sigma$ is defined by (\ref{eq:LSCV-CovarianceMatrix})):

\begin{align}\label{eq:LSCV-H-start}
 H_{start}=(4/(d+2))^{1/(d+4)}n^{-1/(d+4)}\Sigma^{1/2}.
\end{align}

\end{enumerate}

\subsection{Some formula modifications}\label{sec:math-modif}

The equations for LSCV\_h algorithm may be reformulated to require less operations. Let us consider equation \eqref{eq:LSCV-Kx}:

$$K(u)=(2\pi)^{-d/2}|\Sigma|^{-1/2} exp\left( -\frac{1}{2}u^T\Sigma^{-1} u\right).$$

As can be determined from equation \eqref{eq:LSCV-gh}, the $u$ is always equal to $\frac{X_i-X_j}{h}$. Let us therefore reformulate the equation \eqref{eq:LSCV-Kx}:

\begin{equation}
\label{eq:LSCV_MOD-Kv1}
\begin{split}
\tilde{K}(v) & = K(v/h) =  \\
           & = (2\pi)^{-d/2} |\Sigma|^{-1/2} exp \left( -\frac{1}{2}\frac{v}{h}^T\Sigma^{-1}\frac{v}{h}\right) = \\
           & = (2\pi)^{-d/2} |\Sigma|^{-1/2} exp \left( -\frac{1}{2}\frac{1}{h^2}v^T\Sigma^{-1}v\right).
\end{split}
\end{equation}

Let 

\begin{eqnarray}
\label{eq:LSCV_MOD-Pv}
 S(v)=v^T\Sigma^{-1}v.
\end{eqnarray}

If we substitute this into equation \eqref{eq:LSCV_MOD-Kv1} we obtain:

\begin{eqnarray}
\label{eq:LSCV_MOD-Kv2}
\tilde{K}(v) & = & (2\pi)^{-d/2} |\Sigma|^{-1/2} exp \left( -\frac{1}{2}\frac{1}{h^2}S(v)\right).     
\end{eqnarray}

Analogous changes can be made to equation \eqref{eq:LSCV-KKx}:
\begin{eqnarray}
 \label{eq:LSCV_MOD-KKv}
 (\tilde{K}*\tilde{K})(v) &=& (4\pi)^{-d/2} |\Sigma|^{-1/2} exp \left( -\frac{1}{4}\frac{1}{h^2}S(v)\right).  
\end{eqnarray}

These changes can be next propagated to equations \eqref{eq:LSCV-gh} and \eqref{eq:LSCV-Tx}:

\begin{eqnarray}
 \label{eq:LSCV_MOD-Tv}
 \tilde{T}(v) & = & T\left(\frac{v}{h}\right)=(\tilde{K}*\tilde{K})(v) -2 \tilde{K}(v)
\end{eqnarray}

\begin{eqnarray}
\label{eq:LSCV_MOD-gh}
g(h)=h^{-d}\left[2n^{-2}\sum^n_{i=1}\sum^n_{j=1,i<j}\tilde{T}\left( X_i-X_j \right)+n^{-1}R(K) \right]
\end{eqnarray}

\noindent It is easy to notice that $S(v)$ values are scalars, and moreover they are constant, independent on the value of parameter $h$ of the function $g(h)$. Consequently, they may be precalculated for each combination of two vectors X at the start of the algorithm and used multiple times during the search for minimum of $g(h)$. Let us determine the complexity of calculating the $g(h)$ function before and after modifications. Calculating of a single exponent value in either $K(u)$ or $(K*K)(u)$ function requires $O(d^2)$ operations (see section \ref{sec:TwoVectorFun}). These functions need to be calculated $O(n^2)$ times, which leads to the complexity of $O(n^2 d^2)$. Let $n_h$ be the number of times $g(h)$ function needs to be calculated during searching for its minimum. Total complexity of unmodified LSCV\_h algorithm is therefore $O(n_h n^2 d^2)$

Precalculating of a single $S(v)$ value requires $O(d^2)$ operations. $n(n-1)/2$ of $S(v)$ values need to be precomputed. Consequently precomputing of all $S(v)$ values has a complexity of $O(n^2 d^2)$. However, since $S(v)$ values may be reused, computing of the $g(h)$ value has only the complexity of $O(n^2)$. Consequently, total complexity of the modified LSCV\_h algorithm is $O(n^2(d^2+n_h))$. 

The solutions described above unfortunately cannot be used for optimizing of the LSCV\_H algorithm. This is due to the fact that the expression $(X_i-X_j)^T H^{-1}(X_i-X_j)$ found in equations \eqref{eq:LSCV_MULTI-Kx} and \eqref{eq:LSCV_MULTI-KKx} (which is an equivalent to $S(v)$) depends on the $g(H)$ (equation \eqref{eq:LSCV_MULTI-gh}) function argument. Consequently, this expression must be recomputed each time the $g(H)$ function is computed.

\section{Optimization of bandwith selection on SIMD architectures}\label{sec:simd}

In this section we describe parallel algorithms for bandwidth selection and implementation details for two out of three implementations compared in this paper: SSE implementation and GPU implementation. Sequential implementation will be described in section \ref{sec:c-fw}. 

\subsection{Identification of some compute-intensive parts in mathematical formulas}\label{sec:identification}

Let us take a closer look at equations \eqref{eq:PLUGIN-VarianceEstimator}, \eqref{eq:PLUGIN-Psi6Estimate}, \eqref{eq:PLUGIN-Psi4Estimate}, \eqref{eq:LSCV_MULTI-gh} and \eqref{eq:LSCV_MOD-gh}. All of these equations (among other operations) compute sums of large number of values. As such sums are performed multiple times and constitute a large part of the number of basic matematical operations computed in these equations. Accelerating them would significantly increase algorithm performance. Consequently, in general we need an algorithm which given a single row matrix $A$, would compute:
\begin{equation}R(A)=\sum_{i=1}^n A_i.\end{equation}                                      
\noindent The process of using the same operation multiple times on an array of values to obtain a single value (sum, multiplication, mininimum, maximum, variance, count, average etc.) is called \emph{reduction of an array}. Parallel reduction of large arrays is a known problem \cite{XI1998,reduction}. Basic algorithm as well as GPU and SSE implementations, are described in section~\ref{sec:computingSums}.

Let us consider the equation \eqref{eq:PLUGIN-VarianceEstimator}. This equation contains two sums. One of these sums is a sum of values of a scalar function computed based on values stored in a matrix. Formally, given a single row matrix $A$ and a function $fun(x)$, this sum is equivalent to:
\begin{equation}R_{fun}(A)=\sum_{i=1}^n fun(A_i).\end{equation}
\noindent Parallel computation of such sums can be performed by using a simple modification of parallel reduction algorithms. For details refer to section~\ref{sec:computingSumsFunc}. 

Let us now consider equations \eqref{eq:PLUGIN-Psi6Estimate} and \eqref{eq:PLUGIN-Psi4Estimate}. Given a single row matrix $A$ of size $n$ and a function $fun(x)$, both of these equations contain double sums of function values equivalent to:
\begin{equation}RR_{fun}(A)=\sum_{i=1}^n\sum_{j=1,i<j}^n fun(A_i-A_j).\end{equation}
\noindent As can be easily noticed, the function $fun(x)$ is computed for a difference between every combination of values from row matrix $A$. Parallel algorithms for computing such sums are given in section~\ref{sec:sumOf2VarFunc}.

Similar sums can also be found in equations \eqref{eq:LSCV_MULTI-gh} and \eqref{eq:LSCV_MOD-gh}. These sums, given a two dimensional $A$ matrix, and function $fun(x)$ are equivalent to:
\begin{equation}RR^v_{fun}(A)=\sum_{i=1}^n\sum_{j=1,i<j}^n fun(A_{:,i}-A_{:,j}).\footnote{The $v$ superscript stands for \emph{vector}.}\end{equation}
In these equations however, each argument of the function $fun(x)$ is a vector and computation of this function is much more complex. Moreover, in both cases function $fun(x)$ can be expressed as: $fun(x)=fun1(fun2(x))$ where 
\begin{equation}
\label{eq:fun2}
 fun2(x)=x^TMx,
\end{equation}
\noindent $M$ is any matrix and $fun1(y)$ is any scalar function. Let us now consider equation \eqref{eq:LSCV_MOD-gh}. Here, the function $fun(x)$ is an equivalent of the function $\tilde{T}(v)$ presented in equation \eqref{eq:LSCV_MOD-Tv}. Function $\tilde{T}(v)$ is computed using functions $\tilde{K}(v)$ (equation \eqref{eq:LSCV_MOD-Kv2}) and $(\tilde{K}*\tilde{K})(v)$ (equation \eqref{eq:LSCV_MOD-KKv}). These functions in turn can be computed based on a value of the function $S(v)$ (equation \eqref{eq:LSCV_MOD-Pv}) which is an equivalent of $fun2(x)$. As was mentioned earlier in section \ref{sec:math-modif}, the $S(v)$ values can be precomputed and used each time equation \eqref{eq:LSCV_MOD-gh} is computed. We can therefore split the problem of computing the sums in equation \eqref{eq:LSCV_MOD-gh} into two problems: (a) computing $fun2(x)$ ($S(v)$) values (see section \ref{sec:TwoVectorFun}) and (b) finding a sum of values of a scalar function introduced earlier. Section \ref{sec:alg_lscv_h} presents details on how to use precomputed $S(v)$ values in LSCV\_h algorithm. Similar observations can be also made for equation \eqref{eq:LSCV_MULTI-gh}. Here, the function $fun(x)$ is an equivalent of the function $T_H(X_i-X_j)$ presented in equation \eqref{eq:LSCV_MULTI-Tx}. Function $T_H(X_i-X_j)$ is computed using functions $K_H(X_i-X_j)$ (equation \eqref{eq:LSCV_MULTI-Kx}) and $(K*K)_H(X_i-X_j)$ (equation \eqref{eq:LSCV_MULTI-KKx}). Exponents of both functions $K_H(X_i-X_j)$ and $(K*K)_H(X_i-X_j)$ contain $(X_i-X_j)^TH^{-1}(X_i-X_j)$ which is an equivalent of $fun2(x)$. Unfortunately here values of $fun2$ cannot be precomputed as matrix $H^{-1}$ is different every time equation \eqref{eq:LSCV_MULTI-gh} is computed. Nonetheless solutions presented in section \ref{sec:TwoVectorFun} can also be used in this case. Section \ref{sec:alg_lscv_H} presents how to efficiently compute equation \eqref{eq:LSCV_MULTI-gh}.

\subsection{Parallel reduction of array values $R(A)$} \label{sec:computingSums}
\subsubsection{Basic algorithm}
\begin{figure}[t]
 \begin{center}
  \includegraphics[width=7cm]{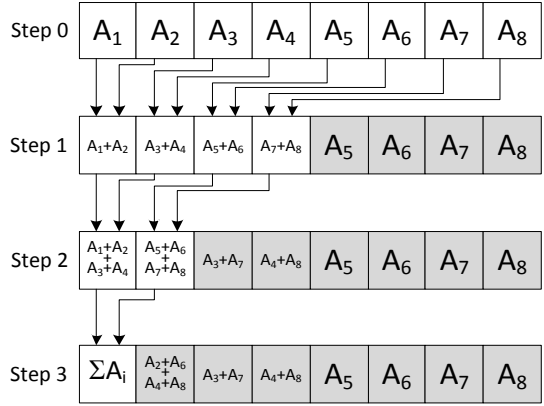}
 \end{center}
 \caption{Parallel reduction schema}\label{fig:reduction1}
\end{figure}

In this section we briefly introduce basic ideas behind the well known problem of parallel reduction of array values \cite{XI1998}, i.e. given a single row matrix $A$,  we present an algorithm which computes
$$R(A)=\sum_{i=1}^n A_i.$$
Basic idea behind the parallel reduction of arrays is presented in Figure \ref{fig:reduction1}. At the beginning the array contains 8 different values. As a first step, pairs of neighbor values from the input array should be added in parallel. Results are stored in the first half of the array. This process is repeated but each time it runs only on the first half of the array which was the input to the previous step. The algorithm stops when the input part of an array is composed on only a single value, which is the result of the parallel reduction. This approach may be easily generalized to larger arrays of values, even of non power-of-two sizes. In such a case, values whose ``pairs'' land outside of array bounds, are left unchanged. The above generic method for value reduction will be adapted in the following sections to accelerate several specific parts of the equations.

\subsubsection{GPU implementation}\label{sec:computingSums:gpu}

GPU based algorithm for reduction of a large number of values was proposed in \cite{reduction} where the author introduces several low-level optimizations which include:
\begin{itemize}
 \item utilizing a small but very fast shared memory available on GPUs (see section \ref{sec:gpu}),
 \item using alternative reduction scheme (pairs of added values are not neighbours in the array),
 \item unrolling of loops and utilizing templates to achieve very efficient gpu-kernel for reduction of arrays.  
\end{itemize}

Paper \cite{reduction} presents a gpu-kernel in CUDA C code, which performs reduction of up to 1024 values (using 512 threads) stored in the shared memory array. Graphics cards with CC$\geq$2.0 may start up to 1024 threads per block, so this code may also be appropriately extended to support arrays of size up to 2048 values. The same paper also introduces a schema which utilizes this gpu-kernel to support arbitrary sized arrays:

\begin{enumerate}
 \item Divide the input dataset into subsets of consecutive values of size equal to some power of 2 (1024 or 2048 depending on compute capability).
 \item Start a one dimensional block of threads for each of these subsets (in a one dimensional grid). Each block should be composed of the number of threads equal to the half of the size of the subset.
 \item Each thread in each block should retrieve two values from the input array (which is initially stored in the slow global memory), add them and store the result in the fast, shared memory. If for some reason grid of threads is not big enough to process all data, each thread should add more then two values.\label{list:includefunctionshere}
 \item Next, the threads should perform the parallel reduction algorithm on the values stored in the shared memory.
 \item The resulting value (for each block) should be stored in the output array in the global memory.
 \item The above steps should be repeated on the output array of the previous iteration until only a single value is obtained. \label{list:additionalreduction}
\end{enumerate}

\subsubsection{SSE implementation}\label{sec:computingSums:sse}
Our SSE implementation uses a horizontal addition instrinsic instruction \texttt{\_\_m128 \_mm\_hadd\_ps(\_\_m128 x,\_\_m128 y)} where \texttt{x} and \texttt{y} are two 128 bit vectors of four floats. Given two vectors \texttt{a=[a1,a2,a3,a4]} and \texttt{b=[a5,a6,a7,a8]} the instruction computes one vector containing \texttt{[a1+a2,a3+a4,a5+a6,a7+a8]}, i.e. it adds neighbour values. As may be easily noticed, the \texttt{\_\_m128 \_mm\_hadd\_ps(\_\_m128 x,\_\_m128 y)} instruction performs first iteration of reduction schema presented in Figure \ref{fig:reduction1}. For bigger arrays, each iteration, except the last two (reducing four values into one), may be performed by executing the instruction on consecutive elements of the array appropriate number of times. 

The algorithm described above requires many memory accesses and may therefore suffer delays stemming from memory access latency. The problem would be largely reduced if the whole temporary array would fit within the CPU's cache. To increase probability of such situation, the processed array is split into many chunks of arbitrary small (but power-of-two) sizes, such that would fit in the cache. Each such chunk is processed independently and the results are then reduced using the same algorithm (similarly as in the GPU approach presented above). 

Notice that the described algorithm has a very low parallelism when compared with the GPU approach (only 4 values are added at the same time). To work around this, we start a number of independent threads, each of which processes a subset of chunks of the processed array. The number of threads dependents on the number of CPU cores and core capability (such as HyperThreading \cite{MBH2002}). 

Careful reader might notice that SSE implementation could be implemented in a much simpler way by using a simple sequential reduction and standard vector addition intrinsic instruction \texttt{\_\_m128 \_mm\_add\_ps(\_\_m128 x,\_\_m128 y)} where \texttt{x} and \texttt{y} are two 128 bit vectors of four floats. Moreover, by using this method memory access latency could be largely reduced. However, as with all of floating point mathematical computations, reduction algorithms are a subject to numerical rounding errors. Sequential reduction has an error constant of $O(n)$, whereas the pairwise reduction algorithm presented above has an error constant $O(log_2 n)$ \cite{H93} and consequently yields smaller errors. One could also argue, that there exists another sequential reduction algorithm introduced by Kahan in \cite{K65}, which has an error constant of $O(1)$ \cite{H93}. This algorithm however requires several times more floating operations and is therefore much slower then simple sequential reduction and pairwise reduction. 

\subsection{Parallel reduction of scalar function values $R_{fun}(A)$}\label{sec:computingSumsFunc}

\subsubsection{Basic algorithm}
In this section we present an algorithm for computing sums equivalent to:
$$R_{fun}(A)=\sum_{i=1}^n fun(A_i),$$
where $A$ is any single row matrix and $fun$ is any scalar function. Notice, that the function $fun$ for each iteration is computed independently. Therefore, values of this function may be easily computed in parallel and the obtained values can be reduced (summed up) later using the previously introduced reduction algorithms (MapReduce approach \cite{DG2008}). However, to remove the need for storing function values in memory we suggest to slightly modify the basic array reduction schema described in section \ref{sec:computingSums} in such a way that $fun$ function values are computed and added on-the-fly. 

\subsubsection{GPU implementation}\label{sec:computingSumsFunc:gpu}
GPU implementation is a straightforward modification of the scheme presented in section \ref{sec:computingSums:gpu}. In step \ref{list:includefunctionshere}, after the values are retrieved from the global memory, $fun$ function values are computed. Afterwards, the computed values are added and stored in the shared memory. This is implemented as a separate gpu-kernel, as subsequent reduction (see step \ref{list:additionalreduction}) is performed without computing of $fun$ function values.

\subsubsection{SSE implementation}\label{sec:computingSumsFunc:sse}
Presence of $fun$ function in SSE implementation of reduction algorithm requires minor adjustments in the basic reduction function. First, a SIMD version of the $fun$ function needs to be implemented. This new version should process four values passed in a single \texttt{\_\_m128} vector, and return four values also as a single \texttt{\_\_m128} vector. The only modification to the SSE reduction implementation requires that in the first iteration of the algorithm, the SIMD version of the $fun$ function is used on every four value vector retrieved from memory and obtained result is subsequently used in the reduction algorithm. 

\subsection{Parallel reduction of scalar function in nested sums $RR_{fun}(A)$} \label{sec:sumOf2VarFunc}
\subsubsection{Basic algorithm}
\begin{figure}[t]
 \begin{center}
  \includegraphics[width=7cm]{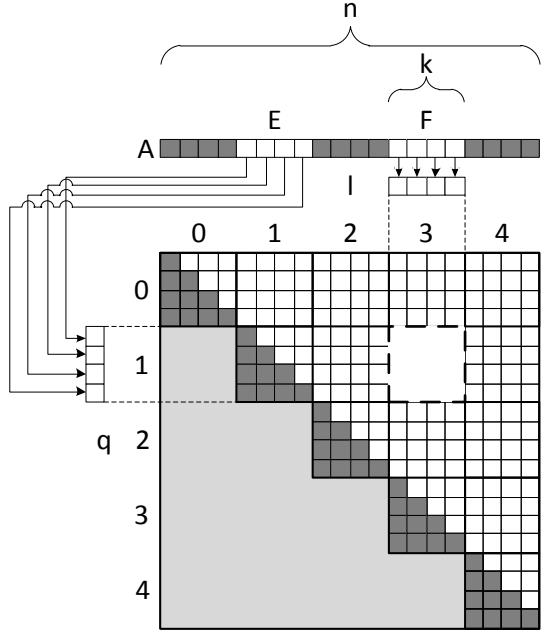}
 \end{center}
 \caption{Parallel cache aware computation of values of two variable functions}\label{fig:matrix_scal}
\end{figure}

In this section we present an algorithm for computing sums equivalent to:
$$RR_{fun}(A)=\sum_{i=1}^n\sum_{j=1,i<j}^n fun(A_i-A_j),$$
where $A$ is any single row matrix and $fun$ is any scalar function. Based on the solutions presented in  \cite{cudaprguide}, we propose the following parallel schema for computing function values, which utilizes cache memory. First, let us notice that $fun$ function values are computed only for indexes $i$ and $j$ such that $i<j$ and might therefore be stored in an upper triangular matrix of size $n \times n$. Lets take a look at Figure \ref{fig:matrix_scal}. The $A$ matrix is divided into chunks of small (power-of-two) size $k$ (recall that here  $A$ matrix contains only a single row). The triangular matrix of $fun$ function values is divided into tiles of size $k\times k$. Notice, that tiles on the diagonal contain redundant positions which should not be computed. Each tile corresponds to some combination of two $A$ matrix chunks denoted $E$ and $F$. For each tile, a group of $k\times k$ threads is started. First, a subset of threads in a group copies the corresponding chunks into the cache memory. Next, each thread in the tile computes the function value based on two  $A$ matrix  values retrieved from the cache. Redundant threads (below the main diagonal) return function value 0. Next, threads in each tile cooperate to reduce the computed function values into a single value using parallel reduction algorithm introduced earlier. Consequently, each tile yields one reduced value. These values are then formed into a single result, by once again using a parallel reduction algorithm. 

The remaining problem is: what to do if the size of the $A$ matrix cannot be expressed as a multiple of the chosen $k$ value? In such a case, the matrix should be extended to the size equal to the nearest multiple of $k$. Redundant threads allocated to process tiles corresponding to last chunk of the array storing $A$ matrix should output zero as a function value. No additional changes to the above schema are required. 

\subsubsection{GPU implementation}\label{sec:sumOf2VarFunc:gpu}

\begin{figure}
 \begin{center}
  \includegraphics[height=5cm]{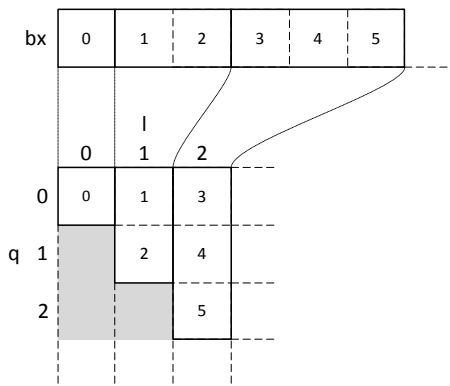}
 \end{center}
 \caption{Conversion of block number to its position in upper triangular matrix}\label{fig:bx2lq}
\end{figure}

The proposed parallel schema seems to fit the GPU architecture and CUDA API pretty well. Each tile can be processed by a square block of threads and the shared memory can be used as a cache. Unfortunately, the computation grids in CUDA API can only be linear or rectangular (no triangular grids). To solve this problem we propose to run a one dimensional grid of blocks, and based on the block number, compute its position in the triangular matrix (see Figure \ref{fig:bx2lq}). To find the position of the tile in the triangle matrix based on its number on the one dimensional grid, we propose to use the equations \eqref{eq:colInTriangleMatrix} and \eqref{eq:rowInTriangleMatrix} (see appendix \ref{sec:jqderiv} for derivation of these equations), where $bx$ is the block number, $l$ is the corresponding tile column and $q$ is the corresponding tile row:

\begin{equation}
\label{eq:colInTriangleMatrix}
l=\left\lceil\frac{\sqrt{8bx+9}-3}{2} \right\rceil,
\end{equation}

\begin{equation}
\label{eq:rowInTriangleMatrix}
q=bx-\frac{l(l+1)}{2}.
\end{equation}

Notice, that as the number of blocks in one dimensional grid cannot exceed 65535 (current GPU limitation), the number of tiles in one dimension of the triangle matrix cannot exceed 360. This in turn limits the number of values in the $A$ matrix  to only $360k$ columns. To solve this problem we allocate two dimensional grids which are composed of at least the appropriate number of blocks and then we find the ``one dimensional'' number of the block based on its position in the two dimensional grid. This new position is used in the subsequent computations in equations  \eqref{eq:colInTriangleMatrix} and \eqref{eq:rowInTriangleMatrix}. The number of blocks started this way may be greater than required. Therefore each thread must detect whether it belongs to one of such redundant blocks and abort computations if necessary.

Given NVIDIA graphics card limitations, $k=16$ (256 thread blocks) could be used for graphics cards with CC$\leq$1.3 and $k=32$ (1024 thread blocks) for graphics cards with CC$\geq$2.0. However, as was observed in \cite{reduction}, given an array of $n$ values, only $n/2$ threads are used during reduction. Therefore, after computing of function values in a tile, only half of the threads in each block would be used in subsequent tile values reduction. To solve this problem we propose to use $k=32$ regardless of graphics cards compute capability, but use blocks of only 256 threads to process tiles. Each thread should compute four $fun$ function values and add them. This way we allow our code to be run on all graphics cards and achieve better thread utilization at the same time.

\subsubsection{SSE implementation}\label{sec:sumOf2VarFunc:sse}

\begin{figure}[t]
 \begin{center}
  \includegraphics[width=8cm]{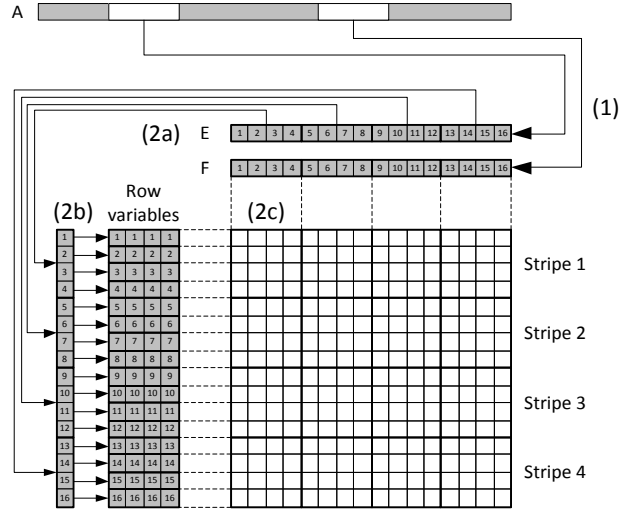}
 \end{center}
 \caption{Processing of a tile in SSE implementation}\label{fig:ssetile}
\end{figure}

In our SSE implementation each tile is processed by a single thread, though some processing parallelism is achieved due to the usage of SSE SIMD instructions (similarly as in section \ref{sec:computingSumsFunc:sse} SIMD version of the function $fun$ is needed). Here, cache is implemented as two local arrays $E$ and $F$ (corresponding to chunks $E$ and $F$) which are read multiple times during computing of tile values and therefore are probably cached. $A$ matrix is divided into 16 value chunks ($k=16$) and consequently each tile is composed of 256 values. However, each such chunk is divided into 4 four value vectors stored in SSE $\_\_m128$ variables. Processing of each tile is illustrated in Figure \ref{fig:ssetile} and described below: 

\begin{enumerate}
 \item First, the two chunks corresponding to a tile are copied to local arrays $E$ and $F$.
 \item Next, the tile is processed in four stripes. Each stripe is processed as follows:
  \begin{enumerate}
   \item A four value vector is retrieved from the $E$ local array which contains ``row values''. 
   \item Next, these four values are copied into four \texttt{\_\_m128} variables (subsequently called row variables).
   \item Each row of a stripe is processed in four steps, each of which includes the following operations: (1) retrieve a four value vector from $F$ array, (2) compute a difference between appropriate row variable and the retrieved vector using \texttt{\_\_m128 \_mm\_sub\_ps(\_\_m128 x, \_\_m128 y)} instruction, (3) compute four $fun$ function values  using SIMD version of the $fun$ function and (4) add the obtained results into a single accumulator \texttt{\_\_m128} type variable using \texttt{\_\_m128 \_mm\_add\_ps(\_\_m128 x, \_\_m128~y)} instruction.
  \end{enumerate}
 \item After all stripes are processed, the accumulator contains four values. These values are added to obtain final result of tile processing. 
\end{enumerate}

Processing of a tile that lies on main diagonal is similar, but several modifications are made. First, only one chunk is retrieved into local array (obviously). Moreover, loops are altered to not process parts of stripes below the main diagonal. Notice however, that some ``below diagonal'' values in small $4\times 4$ tiles are still computed. In such cases, each vector computed by the reduced function is multiplied by appropriate vectors composed of zeroes and ones to reset the unwanted function results. 

The tiles are processed concurrently on a number of threads which are processed most efficiently on a CPU (depending on the number of cores and whether the HyperThreading \cite{MBH2002} capability is available or not).

\subsection{Computing of $fun2$ function values in $RR^v_{fun}(A)$}\label{sec:TwoVectorFun}

\subsubsection{Basic algorithm}\label{sec:TwoVectorFun:alg}
Let us recall that in equations \eqref{eq:LSCV_MULTI-gh} and \eqref{eq:LSCV_MOD-gh} the sums equivalent to:
$$RR^v_{fun}(A)=\sum_{i=1}^n\sum_{j=1,i<j}^n fun(A_{:,i}-A_{:,j})=\sum_{i=1}^n\sum_{j=1,i<j}^n fun1(fun2(A_{:,i}-A_{:,j}))$$ 
\noindent can be found, where $A$ is any matrix, $fun1$ is any scalar function and $fun2(x)=x^TMx$ where  $M$ is also any matrix. As was suggested in section \ref{sec:identification} parallel processing of such sums can be split into two problems: computing $fun2$ function values and then reducing them using previously introduced algorithm. Consequently, we need an algorithm which given an $A$ matrix  would find a triangular matrix $B=[b_{i,j}]$, $i=1,\ldots,n$, $j=i+1,\ldots,n$, such that:

$$b_{i,j}=fun2(A_{:,i}-A_{:,j}).$$
 
\begin{figure}[t]
 \begin{center}
  \includegraphics[width=8cm]{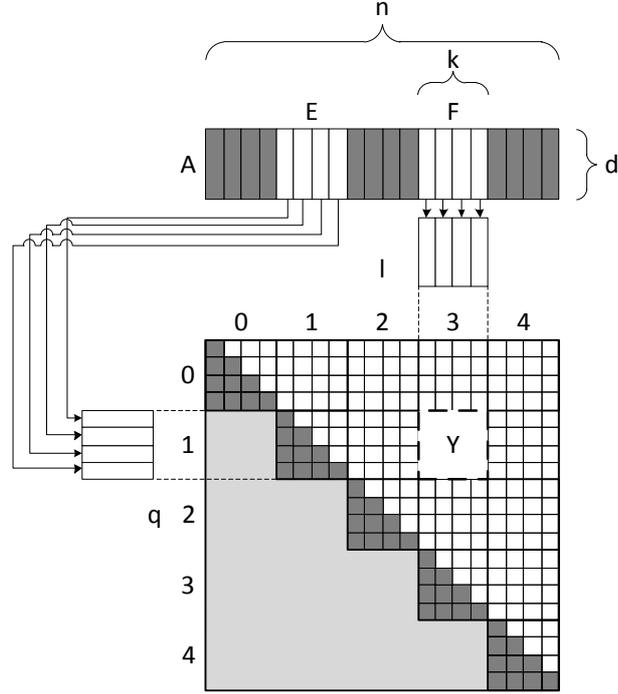}
 \end{center}
 \caption{Parallel computation of $fun2$ function values}\label{fig:matrix_vec}
\end{figure}

\noindent As each $fun2$ function value may be computed independently, parallel computation algorithm seems obvious. To achieve better performance a cache aware solution similar to the one presented in section \ref{sec:sumOf2VarFunc} may be used. Consider the schema presented in Figure \ref{fig:matrix_vec}. $A$ matrix is divided into chunks of $k$ vertical vectors (columns). The triangular result matrix is divided into tiles of size $k\times k$ (notice, that tiles on the matrix diagonal contain excessive positions). Each tile corresponds to some combination of two $A$ matrix chunks. Row of a tile within the triangular matrix is denoted as $q$ and column is denoted as $l$. For each tile, a group of $k\times k$ threads is started. First, a subset of threads in a block copies the corresponding chunks into the cache memory. Next, each thread in the tile computes the function value based on two vectors retrieved from the cached chunks. Each thread detects whether it is over or on the main diagonal or not. If it is below the diagonal it stops further computations. If it is over the main diagonal it computes $fun2$ function value and stores it in the output array. Linear position in the output array may be computed using equation for sum of arithmetic progression, based on the threads coordinates within the triangular array. Notice that the order of stored values is unimportant, as they are subsequently only arguments for functions whose values are later reduced (summed up).

We shall now derive an efficient order of performing operations needed to compute $fun2$ function values within a tile for an array storing $A$ matrix that is row-major aligned in the computers memory. For simplicity let us assume that the considered tile does not lie on the main diagonal. We shall denote the tile of the matrix containing $fun2$ function values as a $k\times k$ submatrix $Y$. Each such tile corresponds to two $d \times k$ ($d$ is the number of rows in $A$ matrix) chunks $E$ and $F$ of the $A$ matrix. Let us assume, that chunks $E$ and $F$ start at columns $qk+1$ and $lk+1$ respectively.
Let:
\begin{equation}
 \begin{aligned}
  i& = qk+r, \\
  j& = lk+p
 \end{aligned}
\end{equation}
\noindent where $r,p = 1,\ldots,k$. Consequently $A_{:,i}=A_{:,qk+r}=E_{:,r}$ and $A_{:,j}=A_{:,lk+p}=F_{:,p}$.
Let $v^{r,p}=E_{:,r}-F_{:,p}=A_{:,i}-A_{:,j}$ be all of the arguments of the $fun2$ function within a tile.
From the definition of function $fun2$, and the new notations introduced above we know that:
\begin{equation}
\label{eq:TwoVectorFun-yrp}
 y_{r,p}=fun2(v^{r,p})=(v^{r,p})^T M v^{r,p}.
\end{equation}
Let us extract the first matrix multiplication operation from the equation \eqref{eq:TwoVectorFun-yrp}: 
\begin{equation}
\label{eq:TwoVectorFun-zrp1}
z^{r,p}=(v^{r,p})^T M.
\end{equation}
The $z^{r,p}$ value is a horizontal vector of $d$ scalars: 
\begin{equation}
\label{eq:TwoVectorFun-zrp2}
z^{r,p}=\left[z^{r,p}_{a} \right]_{a=1,\ldots,d}
\end{equation}
where each $z^{r,p}_{a}$ value is a result of a dot product between $v^{r,p}$ vector and $a$-th column of the $M$ matrix:
\begin{equation}
\label{eq:TwoVectorFun-zrp3}
z^{r,p}_{a}=(v^{r,p})^T M_{:,a}=\sum_{c=1}^d v_c^{r,p}m_{c,a}. 
\end{equation}
Let us now substitute the equation \eqref{eq:TwoVectorFun-zrp2} into equation \eqref{eq:TwoVectorFun-yrp}:
\begin{equation}
 y_{r,p} =   z^{r,p} v^{r,p} = \left[ z^{r,p}_a \right]_{a=1,\ldots,d} v^{r,p}.
\end{equation}
As $v^{r,p}$ is a vertical vector of $d$ values, the above expression is a dot product of vectors $z^{r,p}$ and $v^{r,p}$:
\begin{equation}
\label{eq:TwoVectorFun-yrp2}
 y_{r,p}  = \left[ z^{r,p}_a \right]_{a=1,\ldots,d} \left[v^{r,p}_a \right]_{a=1,\ldots,d} = \sum _{a=1}^d z^{r,p}_a v^{r,p}_a.
\end{equation}
If we substitute equation \eqref{eq:TwoVectorFun-zrp3} into equation \eqref{eq:TwoVectorFun-yrp2} we obtain:
\begin{equation}
\label{eq:TwoVectorFun-yrp3}
 y_{r,p}= \sum_{a=1}^d\left( \sum^d_{c=1} v^{r,p}_c m_{c,a}\right) v^{r,p}_a.
\end{equation}
Recall that $v_x^{r,p}=e_{x,r}-f_{x,p}$. If we substitute this into equation \eqref{eq:TwoVectorFun-yrp3} we obtain:
\begin{equation}\label{eq:yrpFinal}
 y_{r,p}=\sum _{a=1}^d\left( \sum^d_{c=1} (e_{c,r}-f_{c,p}) m_{c,a}\right) (e_{a,r}-f_{a,p}).
\end{equation}
As each $y_{r,p}$ value is computed independently of other $y_{r,p}$ values, we can extend the above equation to compute the whole row $Y_{r,:}$ of $Y$ matrix. This is accomplished by replacing each occurence of the column number $p$ with the colon which means ``all available values''. To retain correctness of terms that are not dependent on $p$ (such as $e_{c,r}$) we introduce the following notation. By $[x]_k$ we denote a horizontal vector of $k$ values equal to x. All terms that are not dependent on $p$ are converted into horizontal vectors of $k$ values. Consequently, the row $r$ of the tile matrix $Y$ may be computed as follows:
\begin{equation}
 \label{eq:Yfinal}
 Y_{r,:}=\sum _{a=1}^d\left( \sum^d_{c=1} ([e_{c,r}]_k-F_{c,:}) m_{c,a}\right) ([e_{a,r}]_k-F_{a,:})
\end{equation}
Notice, that equation \eqref{eq:Yfinal} expresses a single tile row in terms of either single matrix values ($e_{x,r}$ and $\bar{\sigma}_{c,a}$) or chunk rows ($F_{x,:}$). Let us rewrite the above equation in algorithmic form:

\begin{itemize}
 \item[] For each $r=1,\ldots,k$ perform the following steps:
 \begin{enumerate}
  \item $Y_{r,:} \gets [0]_k$
  \item for each $a=1,\ldots,d$ perform the following steps:
   \begin{enumerate}
    \item $part\gets [0]_k$
    \item for each $c=1,\ldots,d$ perform the following step: 
     \begin{itemize}
      \item[] $part\gets part+m_{c,a}*\left([e_{c,r}]_k-F_{c,:}\right)$
     \end{itemize}
    \item $Y_{r,:}  \gets Y_{r,:} + part*\left([e_{a,r}]_k-F_{a,:}\right) $
   \end{enumerate}
  \item Output $Y_{r,:}$
 \end{enumerate}
\end{itemize}

As we assume row-major order storage of the $A$ matrix, rows $F_{x,:}$ are stored in linear portions of the memory, which allows for efficient accesses. Notice that each access to a chunk row is accompanied by an access to a single value in both $M$ matrix and second chunk ($E$). These accesses to memory unfortunately are not to consecutive memory addresses. Notice however that there are only two such accesses per one large linear access to chunk row. Moreover, as $M$ and $E$ matrices are small, they can be easily fit within the cache memory for faster accesses.

\subsubsection{GPU implementation}\label{sec:TwoVectorFun:gpu}
Our GPU implementation is a straightforward implementation of the parallel schema presented in section \ref{sec:TwoVectorFun:alg}. Each tile is processed by a group of 256 threads ($k=16$). We have also limited lengths of vectors to up to 16 values ($d\leq 16$) to simplify copying of data to shared memory (acting as cache) and subsequent processing (256 threads can copy 16 vectors of 16 values in parallel). As CUDA threads are not processed independently (they are run in 32 thread SIMD groups) and are a subject to memory access restrictions, loops computing $fun2$ function values, as well as global memory to shared memory copying code, have to be carefully designed and data has to be properly layed out in memory in order to avoid serialization of memory accesses. 

\begin{figure}[t]
 \begin{center}
  \includegraphics[width=8cm]{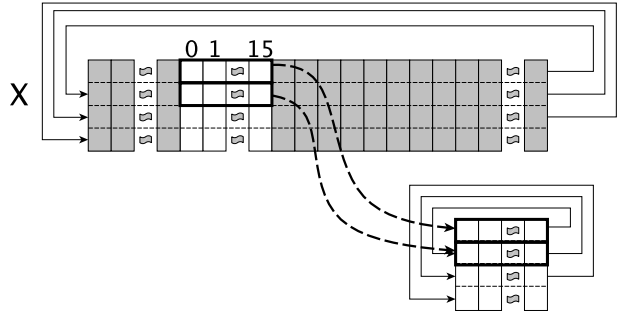}
 \end{center}
 \caption{Copying data to shared memory}\label{fig:memaccess}
\end{figure}

Let us start with copying of data. $A$ matrix is stored in an array in GPUs global memory in row major order. Consequently, subsequent values in the memory belong to subsequent matrix columns. Recall that CUDA threads are grouped in warps. Consider a single memory access instruction. Depending on the compute capability of the graphics card one memory transaction is performed per halfwarp (compute capability $\leq 1.3$) or one per warp (compute capability $\geq 2.0$), as long as each accessed address lies within the same 128B segment within the global memory. On graphics cards with compute capability ($\leq 1.1$) additional restrictions on accessed addresses are imposed. Given the fact that $k=16$, 16 columns should be copied to the shared memory. To adhere to the memory access requirements, threads access consecutive memory addresses (i.e. they copy ``rows'' of the processed chunk of matrix). Consequently, a single thread warp will copy two rows of data. For GPUs with CC$\leq$1.3, a memory transaction per halfwarp is perfomed and consequently, each row will be retrieved in a single 64B memory transaction. For GPUs with CC$\geq$2.0, even though one 128B transaction can be performed per warp, here two 128B transactions will be made, as each row is in different memory segment. Consequently for graphics cards with CC$\geq$2.0, memory retrieval is not as efficient as for the graphics cards with CC$\leq$1.3 (unless we increase $k$ to $32$). However, the excessive retrieved rows will be cached and might benefit other thread blocks. Moreover, this step is short, when compared to next computation steps and the delay should be negligible. Copying of data from global memory to shared memory is depicted in Figure \ref{fig:memaccess}.

To compute $y^{r,p}$ values $M$ matrix is needed as well. Array storing $M$ matrix is stored in row major order in constant memory. Let us assume that the shared memory array storing columns from the chunk corresponding to the ``row'' coordinate of the triangle matrix tile is denoted $E$ whereas the shared memory array storing columns from the chunk corresponding to the ``column'' coordinate of the triangle matrix tile is denoted $F$. By $tx$ and $ty$ we denote thread coordinates within a block. The main loop of a single thread computing $fun2$ function value is as follows:

\begin{enumerate}
 \item $p \gets tx$
 \item $r \gets ty$
 \item $yrp \gets 0$
 \item For each consecutive row $a$
  \begin{enumerate}
   \item $part \gets 0$
   \item For each consecutive row $c$
    \begin{enumerate}
     \item[] $part \gets part + M[c*d+a] * (E[c*k+r]-F[c*k+p])$
    \end{enumerate}
   \item $yrp \gets yrp += part*(E[a*k+r]-F[a*k+p])$
  \end{enumerate}
\end{enumerate}

It is easy to notice that this algorithm is a straightforward implementation of the equation \eqref{eq:yrpFinal}, where the $r$ and $p$ coordinates of a computed value within a tile correspond to threads location within a block. Notice, that each thread in a warp will access the same value in the $M$ array at each iteration of the loop. This is an optimal access pattern for the constant memory where this matrix is stored. 

Let us now analyse the access patterns to the shared memory arrays $E$ and $F$. Data in these arrays is stored in row major order. This is due to the fact that $A$ matrix  is stored in row major order in the global memory and code that retrieves data from $A$ matrix into arrays $E$ and $F$ retains this order. Consequently, first $k$ values in both of these arrays are the values of the first row of the corresponding chunk of the $A$ matrix. Next $k$ values correspond to the second row, and so on. Consequently, consecutive values of each column are stored every $k$ values. Such storage allows our gpu-kernel to access data in the arrays $E$ and $F$ optimally. 

We shall start with an explanation for graphics cards with compute capabilities $\leq 1.3$. Let us assume that $k=16$. Consequently, each half-warp processes one row of a block. As conflicts may only appear between threads in a single half-warp, our subsequent analysis will consider only one row of a block. In Table \ref{tab:breakdown_halfwarp} we present for each thread in a halfwarp its corresponding $p$ and $r$ ($tx$ and $ty$), array address which will be accessed in both $E$ and $F$ arrays as well as corresponding shared memory banks. Each consecutive thread is assigned consecutive $tx$ values \cite{cudaprguide}. Given the fact, that there are 16 shared memory banks, each thread in the half-warp accesses array $F$ through a different bank, which in turn guarantees that there are no conflicts. Let us now consider the $E$ array. It would seem, that in each iteration every thread accesses the same bank. However, it not only accesses the same bank, but also the same address. Consequently, the broadcast mechanism of shared memory may be used and all such reads are performed in a single memory transaction within a half warp. 

For graphics cards with compute capability $\geq 2.0$ explanations are similar. The graphics cards with $\geq$2.0 have 32 shared memory banks, and each memory transaction is performed per warp not per half warp. Given this, all of the above explanations are correct if $k=32$. It is interesting however to notice that even if $k=16$, our gpu-kernel is also efficient. Please take a look at Table \ref{tab:breakdown_warp}. Similarly as in Table \ref{tab:breakdown_halfwarp} we present for each thread in a warp its corresponding $p$ and $r$ ($tx$ and $ty$), array address which will be accessed in both $E$ and $F$ arrays as well as corresponding shared memory banks. If $k=16$ than each warp processes two rows of a block. This causes 16 2-way conflicts between threads accessing the same values in the $F$ array (16 pairs of threads access the same bank), and two 16way conflicts between threads accessing the $E$ array (two groups of 16 threads access the same bank). However, it may also be noticed that in all of these conflicts the same values are accessed and may therefore be broadcasted. The graphics cards with CC$\geq$ 2.0, have an improved broadcast mechanism in which all broadcasts are processed in a single shared memory transaction. Consequently, the described conflicts will not cause memory access serialization. 

\begin{table}
\caption{Breakdown of memory accesses within a single halfwarp on GPUs with compute capability $\leq 1.3$ ($k=16$)} 
\label{tab:breakdown_halfwarp}
\begin{center}
 
\begin{tabular}{|r|c c c c c c|}
\hline
 thread in halfwarp & 0 & 1 & 2 & \ldots & 14 & 15 \\
 $tx=p$ & 0 & 1 & 2 &  \ldots & 14 & 15 \\
 $ty=r$ & $g$ & $g$ & $g$ & \ldots & $g$ & $g$ \\
 F address $c*k+p$ & $16c+0$ &  $16c+1$ & $16c+2$ &\ldots & $16c+14$ & $16c+15$ \\
 F bank & 0 &  1 & 2 &\ldots & 14 & 15 \\
 E address $c*k+g$ & $16c+g$ &  $16c+g$ & $16c+g$ &\ldots & $16c+g$ & $16c+g$ \\
 E bank & $g$ &  $g$ & $g$ &\ldots & $g$ & $g$ \\
\hline
\end{tabular} 
\end{center}
\end{table} 

\newlength{\sep}
\setlength{\sep}{4.5pt}

\begin{table}
\caption{Breakdown of memory accesses within a single halfwarp on GPUs with compute capability $\geq 2.0$ ($k=16$)} 
\label{tab:breakdown_warp}
\begin{center}
\begin{tabular}{|r|@{\hspace{\sep}}c @{\hspace{\sep}}c @{\hspace{\sep}}c @{\hspace{\sep}}c @{\hspace{\sep}}c @{\hspace{\sep}}c|}
\hline
 thread in warp & 0 & 1 & 2 & \ldots & 14 & 15 \\
 $tx=p$ & 0 & 1 & 2 &  \ldots & 14 & 15  \\
 $ty=r$ & $g$ & $g$ & $g$ &  \ldots & $g$ & $g$ \\
 F address $c*k+p$ & $16c+0$ &  $16c+1$ & $16c+2$ &\ldots & $16c+14$ & $16c+15$  \\
 F bank & 0 &  1 & 2 &\ldots & 14 & 15  \\
 E address $c*k+r$& $16c+g$ &  $16c+g$ & $16c+g$ &\ldots & $16c+g$ & $16c+g$  \\
 E bank & $g$ &  $g$ & $g$ &\ldots & $g$ & $g$ \\
\hline
 thread in warp  & 16 & 17 & 18 & \ldots & 30 & 31  \\
 $tx=p$ & 0 & 1 & 2 &  \ldots & 14 & 15  \\
 $ty=r$ & $g+1$ & $g+1$ & $g+1$ & \ldots & $g+1$ & $g+1$  \\
 F address $c*k+p$ & $16c+0$ & $16c+1$ & $16c+2$ &\ldots & $16c+14$ & $16c+15$  \\
 F bank & 0 &  1 & 2 &\ldots & 14 & 15 \\
 E address $c*k+r$ & $16c+g+1$ & $16c+g+1$ & $16c+g+1$ &\ldots & $16c+g+1$ & $16c+g+1$  \\
 E bank & $g+1$ &  $g+1$ & $g+1$ &\ldots & $g+1$ & $g+1$  \\
\hline
\end{tabular} 
\end{center}
\end{table} 

\subsubsection{SSE implementation}
SSE implementation is a straightforward implementation of the generic algorithm presented in section \ref{sec:TwoVectorFun:alg}. Let us start with description of processing of the off-main diagonal tiles. Here we also use $k=16$. First, data is copied into local arrays (which represent $E$ and $F$ matrices) which will hopefully be cached. Subsequent code is just an implementation of the solution from section \ref{sec:TwoVectorFun}. One important thing that should be noted is that we use four variables of type \texttt{\_\_m128} to implement vector $Y_{r,:}$ and another four variables to implement vector $part$. Main diagonal tiles are processed similarly by a specialized variant of the algorithm. The main differences are that code part that outputs results, ignores the excessive values and processing of parts of $Y_{r,:}$ and $part$ vectors is omitted, when applicable. 

Utilizing multiple cores, as previously involves running the two variants of the tile processing algorithm in multiple threads.

\section{Algorithm implementations}\label{sec:algorithms}

In this section we describe how to utilize the algorithms presented in section~\ref{sec:simd} to create efficient SSE and GPU implementations of PLUGIN, LSCV\_h and LSCV\_H algorithms.

\subsection{PLUGIN}\label{sec:alg_plugin}

In our PLUGIN algorithm implementation we utilize parallel reduction algorithms presented in sections \ref{sec:computingSums} and \ref{sec:computingSumsFunc} to compute variance estimator (step~1, section \ref{sec:alg_plugin}, equation \eqref{eq:PLUGIN-VarianceEstimator}) and algorithm presented in section \ref{sec:sumOf2VarFunc} for computing sums in steps 5 and 7 (section \ref{sec:alg_plugin}, equations \eqref{eq:PLUGIN-Psi6Estimate} and \eqref{eq:PLUGIN-Psi4Estimate}). Remaining steps (2,3,4,6 and 8) are all simple equations that are inherently sequential and therefore cannot be further optimized. Nonetheless they require very small number of operations and can therefore be performed on CPU in negligible time. 

\subsection{LSCV\_h}\label{sec:alg_lscv_h}
Our implementations of the LSCV\_h algorithm use the modified equations presented in section \ref{sec:math-modif}. Consequently, the algorithm is performed using the following steps:
\begin{enumerate}
 \item Compute covariance matrix of matrix X: $\Sigma$ (see equations \eqref{eq:LSCV-CovarianceMatrix}, \eqref{eq:LSCV-variance} and \eqref{eq:LSCV-covariance}). \label{list:LSCV_MOD-simpleStart}
 \item Compute determinant of the covariance matrix $\Sigma$: $det(\Sigma)$.
 \item Compute inverse of the covariance matrix $\Sigma$: $\Sigma^{-1}$.
 \item Compute the approximate value of the bandwidth (see equation \eqref{eq:LSCV-h0}). 
 \item Determine the range in which we search for a minimum of the objective function $g(h)$ (see equation \eqref{eq:LSCV-Zh0}). \label{list:LSCV_MOD-simpleEnd}
 \item Compute $S(v^{i,j})$ for all $v^{i,j}=X_i-X_j$ such that $i=1,\ldots,n$ and $j=i+1,\ldots,n$ (see equation \eqref{eq:LSCV_MOD-Pv}). 
 \item Search for minimum of $g(h)$ objective function within the range computed previously (step 5). Each time objective  is computed, its modified version (see equation \eqref{eq:LSCV_MOD-gh}) should be used, which can be computed based on precomputed values of the $S(v)$ function. 
\end{enumerate}

Steps \ref{list:LSCV_MOD-simpleStart} to \ref{list:LSCV_MOD-simpleEnd} of the algorithm in all implementations are performed sequentially on CPU without using SSE instructions. Though computing of covariance matrix could be performed easily in parallel by using an algorithm similar to the one presented in section \ref{sec:TwoVectorFun} we have not implemented this as this step takes very little time when compared to the last steps of the LSCV\_h algorithm presented above. Values of $S(v)$ function are precomputed using the algorithm presented in section \ref{sec:TwoVectorFun}. The last step (minimization of $g(h)$ function) is performed by a ``brute force'' search for a minimum on a grid, where the density of a grid is based on a user specified parameter and should be sufficiently dense. Note that as was stated in section \ref{sec:lscv_h} other approaches to minimization of $g(h)$ objective function are also possible. In this paper however we present implementations of the ``brute force'' method.

\subsubsection{GPU implementation}\label{sec:alg_lscv_h:gpu}
In GPU implementation, searching for minimum of $g(h)$ is performed as follows. First, based on the grid density and the width of the range $Z(h_0)$ (equation \eqref{eq:LSCV-Zh0}) the number of different $h$ values to be tested (denoted $n_h$) are determined. Next, the number of one dimensional blocks needed to perform reduction of the nested sums in equation \eqref{eq:LSCV_MOD-gh} (denoted $n_b$) are determined. Given the block size $bs$ this may be computed as: $n_b=\left\lceil (n(n-1)/2)/(2bs) \right\rceil$. Given these values, a two dimensional computing grid is started, where each row corresponds to a single tested $h$ value (consequently there are $n_h$ rows). Each row is composed of $n_b$ one dimensional blocks. Parallel computation of $g(h)$ values for all tested arguments is performed similarly as in algorithm presented in section \ref{sec:computingSumsFunc}. The main differences here are as follows:

\begin{itemize}
 \item Each thread based on the $y$ coordinate of its block within a computation grid computes tested argument $h$ (its not retrieved from the memory).
 \item Reduction is performed on the precomputed $S(v)$ values. For each retrieved $S(v)$ value and based on the grid row dependent $h$ argument value, the function $\tilde{T}(v)$ is computed and these computed values are then added. 
 \item Reduction is performed independently in each row of the computational grid, i.e. each row reduces the same set of $S(v)$ values but for different $h$ argument value. 
\end{itemize}

Similarly as in section \ref{sec:computingSumsFunc} reduction is performed in each block, so each started block yields one reduced value. An $n_h \times n_b$ matrix of values obtained in this way is stored in global memory. As we expect a single value per row, not a single value per block, subsequent reduction is performed in each row independently. This process is repeated until only a single value per grid row is obtained. The obtained values represent nested sums from the equation \eqref{eq:LSCV_MOD-gh}. To find the final $g(h)$ values, for each value obtained during reduction step, a single thread is started, which performs remaining operations of the equation \eqref{eq:LSCV_MOD-gh}. Computed $g(h)$ values are copied to the computers memory and the argument for which the function $g(h)$ has minimal value is found. Notice that the last operation could also be perfomed in parallel on GPU. However, this step can be performed in negligible time so there is no need to accelerate it. 

Now we would like to address a problem that comes from GPU limitations. The computation grid can have no more than 65535 rows and columns. Consequently, if $n_h$ or $n_b$ cross this boundary, the corresponding computation grid cannot be started. Solving the problem with too big $n_b$ value is pretty simple. If $n_b$ is bigger than 65535, only 65535 blocks per grid row are started. The algorithm for reduction of values presented in \cite{reduction} adds more than two values per thread if there are more values to be reduced, than two times the available number of threads. If $n_h$ is bigger than 65535 the set of $h$  values to be tested is divided into portions of size 65535 and the computations described above are performed sequentially for each such portion. 

\subsubsection{SSE implementation}\label{sec:alg_lscv_h:sse}
As SSE implementation has much less threads to utilize, we have used a simpler approach. Similarly, as in GPU implementation we first determine the value $n_h$. In contrast to GPU approach however, we do not compute $g(h)$ function for each $h$ argument in parallel. This is performed sequentially in a loop, and only computation of a single $g(h)$ function is parallelized. To perform nested sums from the equation \eqref{eq:LSCV_MOD-gh}, we use the algorithm presented in section \ref{sec:computingSumsFunc}. This algorithm is started on the precomputed set of $S(v)$ values and for each such value the function $\tilde{T}(v)$ is computed and subsequently reduced into a single value. Based on this value, the final value of $g(h)$ function is computed. The loop keeps track of the lowest $g(h)$ value found up-to-date (and corresponding $h$ value) and after the loop is finished the result of the LSCV\_h algorithm is known. 

\subsection{LSCV\_H}\label{sec:alg_lscv_H}

The LSCV\_H algorithm can use any numerical function minimization algorithm capable of minimizing $g(H)$ function (see equation \eqref{eq:LSCV_MULTI-gh}). Such algorithms are often inherently sequential as they are based on iterative improvement of results of previous iteration. Consequently, the only parallel approach to such algorithms would require to start multiple parallel instances of this algorithm, each starting from a different starting point in hope of finding better result after a fixed number of steps. However, the number of steps needed to converge to a local optimum cannot be reduced. Possibly, for some specific algorithms, some steps could be parallelized, but that is not a subject of this paper. Still, there is one thing that can be improved here. Notice, that an iterative optimization algorithm needs to compute objective function at least once per iteration to assess currently found solution. Our objective  function $g(H)$ can take a long time to compute, and while it is computed, other optimization algorithm steps cannot be processed. Consequently, the time of finding the optimal $H$ matrix can be improved if we optimize computing of $g(H)$ function. Unfortunately, we cannot use a set of precomputed values like the ones used to speed up the LSCV\_h algorithm, but we can adapt the algorithm used to compute those values. Compare exponents in equations \eqref{eq:LSCV_MOD-Kv2} and \eqref{eq:LSCV_MOD-KKv} (algorithm LSCV\_h) with exponents in equations \eqref{eq:LSCV_MULTI-Kx} and \eqref{eq:LSCV_MULTI-KKx} (algorithm LSCV\_H). Both exponents are computed similarly. The main difference is that in  LSCV\_h algorithm the matrix $\Sigma^{-1}$ is constant and exponents may be precomputed, while the corresponding $H^{-1}$ matrix in LSCV\_H algorithm is not constant. Consequently, to compute a single value of $g(H)$ both steps: computing exponents and reducing $T$ function value have to be performed. To make this solution a little bit more cache friendly, we combine both: exponent finding algorithm described in section \ref{sec:TwoVectorFun} and function value reduction algorithm presented in section \ref{sec:computingSumsFunc} into one algorithm. 

\subsubsection{GPU implementation}\label{sec:alg_lscv_H:gpu}
Let us start with describing the modification of the GPU algorithm. Recall the algorithm description from section \ref{sec:TwoVectorFun:gpu}. The processing in this algorithm is done in tiles: 256 threads process 256 combinations of two vectors from the matrix X. Each tile is represented as a single thread block. We adapt this algorithm by adding additional steps for each thread, after it finishes computing of its corresponding value. Based on this value, $T(H)$ function (see equation \eqref{eq:LSCV_MULTI-Tx}) is computed and the result is stored in the additional buffer in the shared memory. Next, all threads within a block are synchronized and then perform parallel reduction algorithm on the obtained values. Consequently, after all threads within a block finish processing a tile, a single value, which constitutes partially performed nested sums from the equation \eqref{eq:LSCV_MULTI-gh} is obtained. This value is then stored in global memory for further reduction by using the algorithm from section \ref{sec:computingSums}.  The computed sum is then used to finalize computation of the $g(H)$ function. 

\subsubsection{SSE implementation}\label{sec:alg_lscv_H:sse}
Modifications of the SSE algorithm are pretty similar. Recall that here each 16 value row processed is implemented by four \texttt{\_\_m128} variables. Consequently, each computed row of a processed tile is returned in this form. After such row is computed, it is not stored in the resulting array, but SIMD implementation of the $T(H)$ function is computed on each of these four variables. Next, all of these variables are added to a single \texttt{\_\_m128} accumulator. This process is repeated for all rows computed during processing of a tile. Finally, the four values stored in the accumulator are added to obtain a single value which constitutes partially performed nested sums from the equation \eqref{eq:LSCV_MULTI-gh}. These values are then stored into an array for further reduction by using the algorithm from section \ref{sec:computingSums}.  Computed sum is then used to finalize computing of the $g(H)$ function. 

\section{Experiments}\label{sec:experiments}
\subsection{Environment and implementation versions} \label{sec:implementations}
For the purpose of experiments we have implemented PLUGIN, LSCV\_h and LSCV\_H algorithms (see section \ref{sec:algorithms}), each in 3 versions: Sequential implementation, SSE implementation and GPU implementation. LSCV\_H implementations only implemented computing of the $g(H)$ objective function, as this is the only element of this algorithm that has influence on its performance. We have also made a small change in LSCV\_h algorithm. To make performance of this algorithm completely data independent, we have slightly modified the last part of the algorithm where we search for a minimum of the objective function on a grid. The objective function is computed for a fixed number of points (150) within the computed interval as opposed to section \ref{sec:lscv_h} where only distance between the tested points is specified, and their number is dependent on the width of the computed interval. In all versions we have used ALGLIB library \cite{alglib} to perform matrix square root and Horner's method for accelerating calculations of polynomial values (in portions of equations \eqref{eq:PLUGIN-Psi6Estimate} and \eqref{eq:PLUGIN-Psi4Estimate}). Each implementation uses single precision arithmetic. As was stated in section \ref{sec:gpu} single precision arithmetic is much more efficient on GPUs than double precision. Notheless, performance of double precision computations on GPUs grow with each new graphics card and should be adequate in the near future. Only simple modifications to the presented solutions, which take into account mainly efficient access to shared memory, need to be made in order to accomodate double precision. To make the results of performance tests of CPU and GPU comparable we have decided that the Sequential and SSE implementations should also utilize single precision. Still one should notice that both of these implementations could utilize double precision as well. For sequential implementations changes are trivial. SSE implementation could be changed in two ways to accomodate double precision: either operate on 128bit vectors holding two double precision instead of four single precision values (utilize SSE2 instructions) or operate on 256bit vectors containing four double precision values (utilize AVX instructions and new registers). Below we give a short description of each of the versions. 

\begin{itemize}
 \item \textbf{Sequential implementation} - a ``pure C'' straightforward sequential implementation of formulas presented in sections \ref{sec:plugin} (PLUGIN algorithm),  \ref{sec:lscv_h} and \ref{sec:math-modif} (LSCV\_h algorithm) algorithm, and \ref{sec:lscv_H} (LSCV\_H algorithm). This implementation does not take into account any knowledge about the environment it is going to be executed in. It is not cache aware. The only cache awarness in this implementation is reflected in loops construction and memory storage which avoids non linear memory accesses (up to transposing matrices if necessary). This implementation does not use any SSE instructions and is single threaded. Reductions are performed only using sequential implementation of the hierarchichal algorithm presented in section \ref{sec:computingSumsFunc}. 

 \item \textbf{GPU implementation} - implementation that accelerates computations using CUDA API to execute computations on a GPU. This implementation is highly parallel and uses thousands of threads. Implementation tries to utilize multiple GPU memory types, including very fast shared memory, which may be treated as a user programmable cache. Implementation also uses C++ templates to automatically create several versions of each gpu-kernel with unrolled loops. For details see section \ref{sec:alg_plugin} (PLUGIN algorithm), \ref{sec:alg_lscv_h:gpu} (LSCV\_h algorithm) and \ref{sec:alg_lscv_H:gpu} (LSCV\_H algorithm). 

 \item \textbf{SSE implementation} -  implementation that tries to utilize all ways of accelerating performance available on CPUs, i.e. it utilizes SSE instructions, multiple threads (to utilize multiple CPU cores and HyperThreading capabilities) and is cache aware. OpenMP \cite{CJP2007} is used for thread management. Implementation also uses C++ templates to automatically create several versions of each function with unrolled loops. For details see section \ref{sec:alg_plugin} (PLUGIN algorithm), \ref{sec:alg_lscv_h:sse} (LSCV\_h algorithm) and \ref{sec:alg_lscv_H:sse} (LSCV\_H algorithm). 
\end{itemize}

Experiments were performed on a computer with Intel Core i7 CPU working at 2.8GHz, NVIDIA GeForce 480GTX graphics card and 8GB of DDR3 RAM. The CPU supports SSE4 instruction set, has 4 cores and HyperThreading capability. The GPU has 15 multiprocessors, each composed of 32 streaming procesors. All implementations were run under Linux operating system (Arch Linux distribution, 3.3.7 kernel version).

All of the tested implementations may be downloaded from \url{http://www.cs.put.poznan.pl/wandrzejewski/_sources/hcode.zip}.

\subsection{Experiments and datasets}
We have performed several experiments testing the influence of the number of samples ($n$) and their dimensionality ($d$) on the performance of all of the implementations introduced in section \ref{sec:implementations}. 
The input sample sizes and dimensionalities for each of the algorithms were as follows:
\begin{itemize}
 \item PLUGIN algorithm: $n=1024, 2048,\ldots,32768$, $d=1$ (PLUGIN algorithm only supports one dimensional data).
 \item LSCV\_h algorithm: $n=64,128,\ldots,1024$, $d=1,\ldots,16$.
 \item LSCV\_H algorithm: $n=1024, 2048,\ldots,16384$, $d=1,\ldots,16$.
\end{itemize}

All processing times measured for GPU implementations include data transfer times between GPU and CPU. 

From the obtained processing times we have calculated speedups achieved for each algorithm and its implementation:
\begin{itemize}
 \item SSE over Sequential implementation,
 \item GPU over Sequential implementation,
 \item GPU over SSE implementation.
\end{itemize}
\noindent The results are presented in the next section.

As the input data does not influence processing times of tested implementations (except maybe LSCV\_H method where the performance of a minimization algorithm depends on the dataset, but in our implementation we skip this part), we supply random dataset for each algorithm/sample size/dimensionality size combination. In other words, each implementation of an algorithm is tested in the same conditions. 

\subsection{Experimental results}

\begin{figure}%
\centering
\subfloat[GPU implementation speedup over Sequential implementation]{\label{fig:PLUGIN_speedup:CUDA}
 \includegraphics[width=9cm]{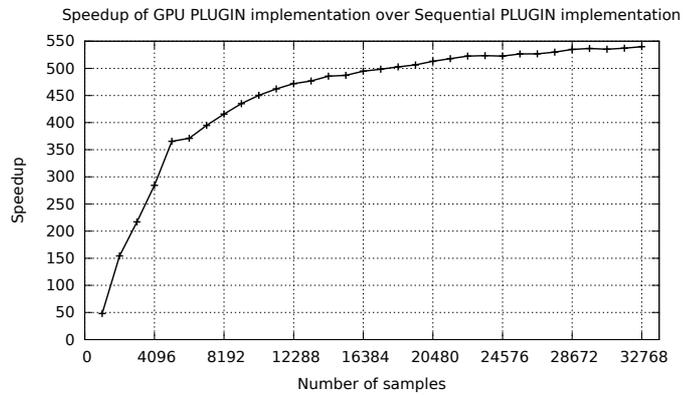}
}%
\qquad
\subfloat[SSE implementation speedup over Sequential implementation]{\label{fig:PLUGIN_speedup:SSE}
 \includegraphics[width=9cm]{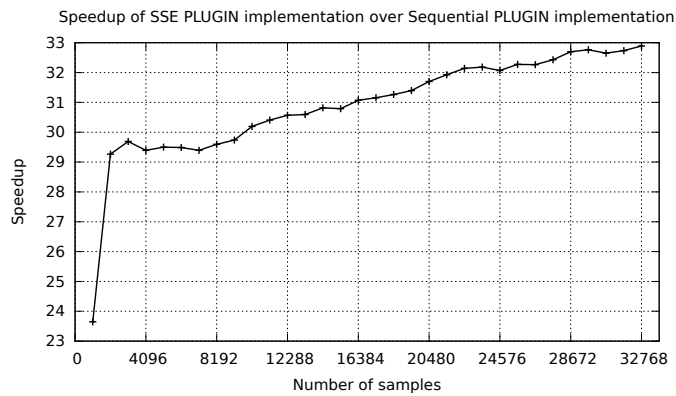}
}%
\qquad
\subfloat[GPU implementation speedup over SSE implementation]{\label{fig:PLUGIN_speedup:CUDASSE}
 \includegraphics[width=9cm]{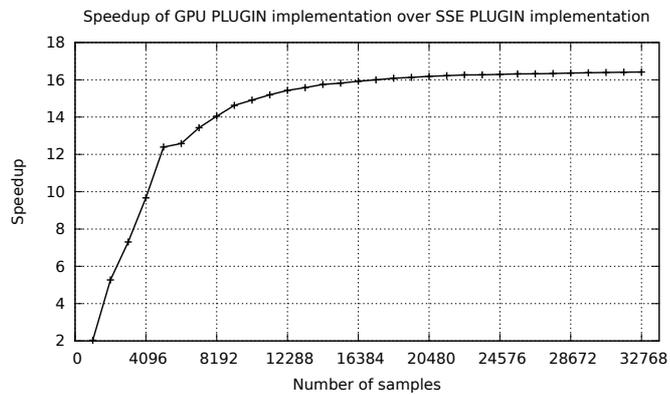}
}%
\qquad
\caption[]{Total speedups of all PLUGIN algorithm implementations}
\label{fig:PLUGIN_speedup}
\end{figure}

Let take a look at Figure \ref{fig:PLUGIN_speedup:CUDA}. It presents speedups of GPU PLUGIN algorithm implementation over Sequential implementation. As can be noticed, GPU implementation is about 500 times faster than sequential implementations. Moreover, notice that the bigger the instance is the greater speedup is obtained. Figure \ref{fig:PLUGIN_speedup:SSE} presents speedups of SSE PLUGIN algorithm implementation over Sequential implementation. The obtained speedups are about 32. Notice, that similarly as with GPU implementation, the obtained speedup grows as the size of the instance increases. Figure \ref{fig:PLUGIN_speedup:CUDASSE} presents speedups of SSE PLUGIN algorithm implementation over Sequential implementation. The maximum obtained speedup is about 16 and the speedup grows as the size of the instance increases. 

First, let us explain the observed increase in obtained speedups. PLUGIN algorithm has the complexity of $O(n^2)$ (the most complex part is computing double sums in equations \eqref{eq:PLUGIN-Psi6Estimate} and \eqref{eq:PLUGIN-Psi4Estimate}). Consequently, for each implementation there exists a second order polynomial which can be used to determine the time needed to process an instance of size $n$. Speedup is calculated by dividing the processing times of the slower implementation by the processing times achieved by the faster implementation. Let $F(n)=a_fn^2+b_fn+c_f$ be the polynomial that determines the processing time of the faster implementation and $S(n)=a_sn^2+b_sn+c_s$ be the polynomial that determines the processing time of the slower implementation. Speedup is calculated as follows:

\begin{equation}
 Speedup(n)=\frac{S(n)}{F(n)}=\frac{a_sn^2+b_sn+c_s}{a_fn^2+b_fn+c_f}
\end{equation}

\noindent Notice that:

\begin{equation}
 lim_{n\rightarrow 0} Speedup(n)=\frac{c_s}{c_f},
\end{equation}

\noindent whereas:

\begin{equation}
 \label{eq:speedupLimit}
 lim_{n\rightarrow \infty} Speedup(n)=\frac{a_s}{a_f}.
\end{equation}

\noindent It can be therefore determined that speedup starts from some value $\frac{c_s}{c_f}$ and asymptotically approaches $\frac{a_s}{a_f}$ for sufficiently big instances. Here $\frac{c_s}{c_f}$ value is smaller than $\frac{a_s}{a_f}$ and consequently the speedup increases with the number of samples $n$. Using this framework we can also determine the maximum speedup in the given hardware and software conditions, by first fitting a second order polynomial to the obtained processing times using least-squares measurement and then calculating limit using equation \eqref{eq:speedupLimit}.

The speedup values shown in Figure \ref{fig:PLUGIN_speedup:SSE} can be explained as follows. The SSE algorithm uses SIMD instructions, performing the same operation on four different values. Moreover, a quad core processor with hyperthreading capabilities is used (sequential implementation uses only a single thread and hence only one core). Four cores times four times faster processing of data should lead to about 16 times speedup. Notice however, that due to HyperThreading capabilities, each core can process multiple threads more efficiently (we have used 8 threads - two per core). Moreover, additional cache aware optimizations make the SSE implementation even faster as little time is wasted on data retrieval from RAM. 

\begin{figure}%
\centering
\subfloat[GPU implementation speedup over Sequential implementation]{\label{fig:LSCV_speedup:CUDA}
 \includegraphics[width=9cm]{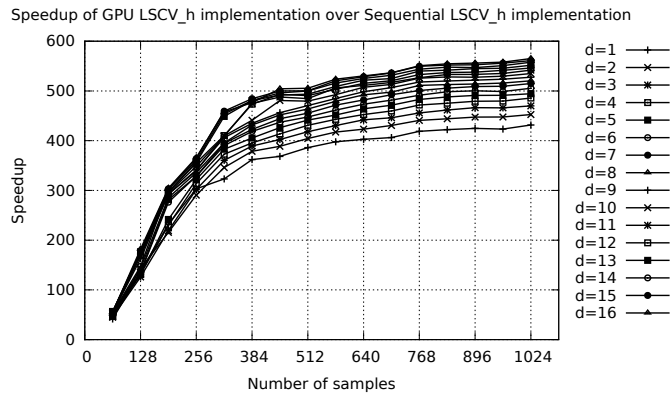}
}%
\qquad
\subfloat[SSE implementation speedup over Sequential implementation]{\label{fig:LSCV_speedup:SSE}
 \includegraphics[width=9cm]{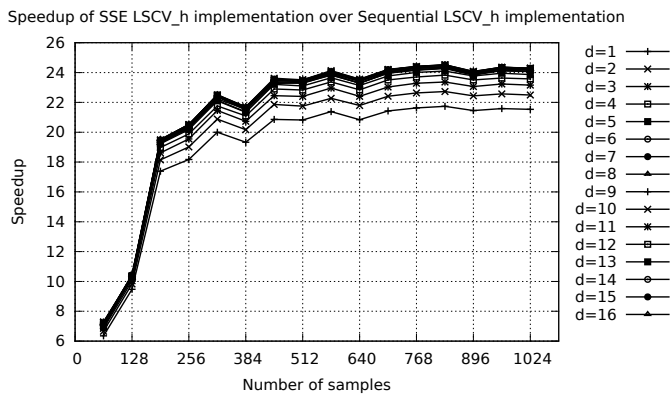}
}%
\qquad
\subfloat[GPU implementation speedup over SSE implementation]{\label{fig:LSCV_speedup:CUDASSE}
 \includegraphics[width=9cm]{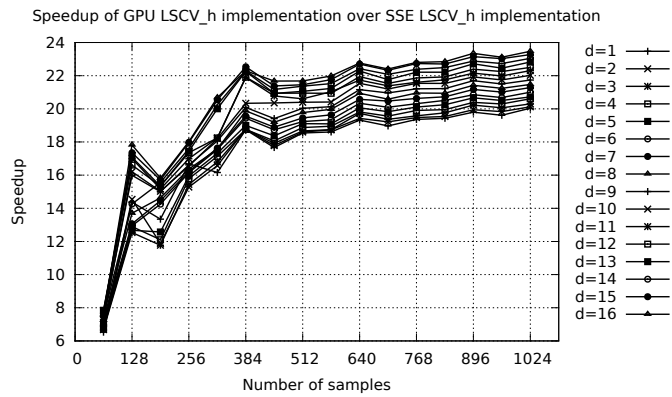}
}%
\qquad
\caption[]{Total speedups of all LSCV\_h algorithm implementations}
\label{fig:LSCV_speedup}
\end{figure}

The speedup of about 16 of GPU implementation over SSE implementation and 500 over sequential is hard to explain similarly as was done for the SSE over Sequential implementation speedup as both architectures are very different.

Lets us now take a look at Figure~\ref{fig:LSCV_speedup}. It presents speedups of GPU LSCV\_h implementation and SSE implementation over Sequential implementation (Figures \ref{fig:LSCV_speedup:CUDA} and \ref{fig:LSCV_speedup:SSE} respectively) and Speedup of GPU implementation over SSE implementation (Figure \ref{fig:LSCV_speedup:CUDASSE}). Each curve represents a different data dimensionality. Speedups achieved here are as follows:
\begin{itemize}
 \item GPU implementation is about 550 times faster than Sequential implementation,
 \item SSE implementation is about 20 times faster than Sequential implementation,
 \item GPU implementation is about 20 times faster than SSE implementation. 
\end{itemize}
\noindent Observations that can be made here are very similar to the ones made for PLUGIN algorithm, as:
\begin{itemize}
 \item Algorithm's complexity is $O(n^2(d^2+n_h))$ (see section \ref{sec:math-modif}). Given the fact that $d$ and $n_h$ are constant, the complexity is reduced to $O(n^2)$, i.e. the same as complexity of the PLUGIN algorithm.
 \item Most of the algorithm processing time constitutes of computing of $g(h)$ function (see equation \eqref{eq:LSCV_MOD-gh}) which is performed by using data reduction algorithms (the same algorithms are used in PLUGIN implementations).
\end{itemize}

One can also make several interesting observations. 

First, for each dimensionality the speedup limit is slightly different. This is due to the compiler optimizations in which loops are unrolled, and for each dimensionality the compiler creates a different version of used procedures. Consequently, each curve represents in fact a slightly different program. Moreover, the bigger the dimensionality, the bigger the speedup is, but for bigger dimensionalities the difference between them is smaller. This may be explained similarly as we have explained a similar fenomenon earlier (the one in which the speedup was increasing with respect to the number of samples $n$). Let us recall that LSCV\_h algorithms complexity is $O(n^2(d^2+n_h))$. For a specified constant values of $n$ and $n_h$ (arbitrarily set to 150), the complexity is reduced to $O(d^2)$. Consequently, processing times may be determined by a second order polynomial where dimensionality is the independent variable. As speedup is calculated by dividing the processing times of the slower implementation by the times achieved by the faster implementation, it is basically a proportion of two second order polynomials, which has a limit at $d\rightarrow\infty$. 

Second observation is that the curves in Figures \ref{fig:LSCV_speedup:SSE} and \ref{fig:LSCV_speedup:CUDASSE} are not smooth. These disturbances are caused by SSE implementation which achieved for $n=128$ and to a lesser extent $n=384$ a little bit worse performance then expected. 

\begin{figure}%
\centering
\subfloat[GPU implementation speedup over Sequential implementation]{\label{fig:LSCV_MULTI_speedup:CUDA}
 \includegraphics[width=9cm]{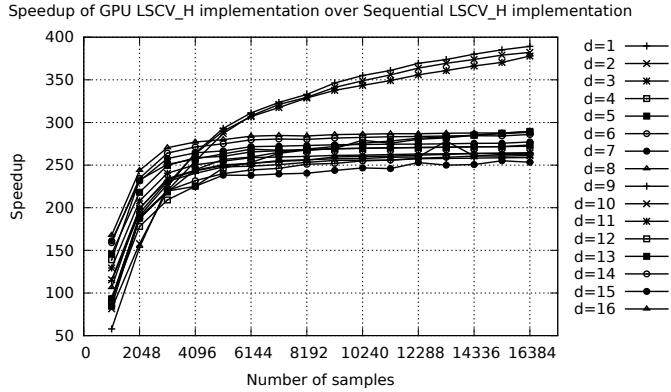}
}%
\qquad
\subfloat[SSE implementation speedup over Sequential implementation]{\label{fig:LSCV_MULTI_speedup:SSE}
 \includegraphics[width=9cm]{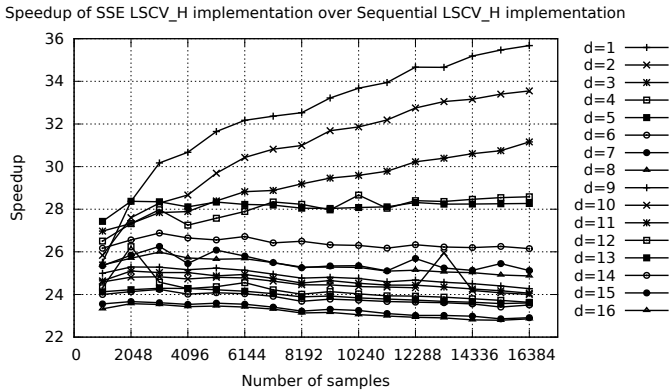}
}%
\qquad
\subfloat[GPU implementation speedup over SSE implementation]{\label{fig:LSCV_MULTI_speedup:CUDASSE}
 \includegraphics[width=9cm]{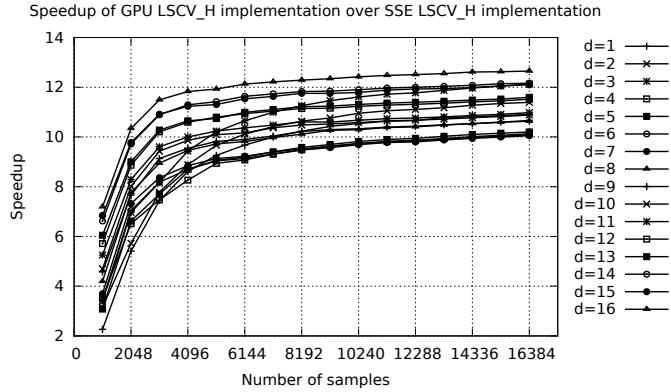}
}%
\qquad
\caption[]{Total speedups of all LSCV\_H algorithm implementations}
\label{fig:LSCV_MULTI_speedup}
\end{figure}

Finally, let us analize Figure \ref{fig:LSCV_MULTI_speedup} which presents speedups of GPU LSCV\_H implementation and SSE implementation over Sequential implementation (Figures \ref{fig:LSCV_MULTI_speedup:CUDA} and \ref{fig:LSCV_MULTI_speedup:SSE} respectively) and speedup of GPU implementation over SSE implementation (Figure \ref{fig:LSCV_MULTI_speedup:CUDASSE}). Similarly, as in LSCV\_h algorithm we have tested processing times for data dimensionalities equal to $d=1,\ldots,16$ and computed speedups based on the obtained values. 
Speedups achieved here are as follows:
\begin{itemize}
 \item GPU implementation is 290 times faster than Sequential implementation,
 \item SSE implementation is 20 times faster than Sequential implementation,
 \item GPU implementation is 10 times faster than SSE implementation. 
\end{itemize}
\noindent Most of the discussion for PLUGIN and LSCV\_h algorithms presented earlier can also be used to explain results obtained for LSCV\_H algorithm implementations. Recall that our LSCV\_H implementation is not complete. We have only tested processing time needed to compute $g(H)$ function (equation \eqref{eq:LSCV_MULTI-gh}). Let us determine the complexity of computing $g(H)$ function. The $g(H)$ function contains double sums which add $O(n^2)$ $T_H(X_i-X_j)$ function values. Computation of the $T_H(X_i-X_j)$ function requires $O(d^2)$ operations. Consequently, computation order of $g(H)$ function is $O(n^2d^2)$. This in turn leads to the earlier discussion of representing function processing times with second order polynomials with either $n$ or $d$ as independent variable, and the second variable constant. As was stated earlier, if processing times are expressed by second order polynomials, then speedup function (defined as ratio of two polynomials) is a function that has a limit at $0$ and at $\infty$. All of the observed values approach some asymptotic value which is equal to the limit of the speedup function at the infinity boundary. Starting value (for low $n$ or $d$ values) is close to the speedup function limit at $n$ (or $d$) tending to $0$. Several other observations can be made. First, take a look at Figure \ref{fig:LSCV_MULTI_speedup:CUDA}. Curves for dimensionalities $d=1,\ldots,3$ are different from the rest of the observed ones. Notice that here the speedup seems to drop with the increase in the number of samples ($n$). These observations however, though look differently than previously analysed on other figures, also fit our ealier discussion regarding limits of the speedup function defined as a ratio between two second order polynomials. Let us consider a situation where $n=16384$. As $n$ is constant, processing times can be determined by a second order polynomials dependent on $d$. For low values of $d$ ($d=1,\ldots,3$) we observe high speedups which descrease as $d$ increases. Such observation means that probably the constant part of the polynomial is smaller for GPU implementation than for Sequential Implementation. However, in such a case it would mean that GPU implementation requires less initialization time than Sequential version. This is certainly not true, as GPU Implementation needs to perform several additional tasks, such as data transfer to device memory. This phenomenon may be explained as follows. Even though the number of dimensions is low, the sequential implementations outer loops are dependant on $n^2$ (they iterate over every combination of two samples from the dataset). This means that there is a constant (we assume $n=16384$) processing time required for processing of those loops (branch prediction etc.). Low values of $d$ mean that the inner loops, which compute equation \eqref{eq:LSCV_MULTI-Tx}, require less time and therefore constitute a lesser percentage of the whole algorithm processing time. The same situation does not influence GPU implementation as much, as outer loop iterations are performed in parallel. Consequently less time is wasted on $n^2$ dependant loop processing costs. This in turn leads to conclusion that for high $n$ and low $d$ speedup is higher. Now, let us assume that $n=1024$. Low values of $n$ mean low outer loop processing costs and GPU typical initialization costs start to dominate, which leads to smaller speedups for low values of $d$. This can be also noticed in Figure \ref{fig:LSCV_MULTI_speedup:CUDA}. One could ask why the same phenomenon was not observed for LSCV\_h algorithm. This stems from the fact that in LSCV\_h algorithm the values of equation \eqref{eq:LSCV_MOD-Tv} (which is equation's \eqref{eq:LSCV_MULTI-Tx} counterpart) are computed only once, and other algorithm's tasks dominate.

Finally, in Table \ref{tab:processingTimes}, for illustrative purposes only, we present comparison of processing times obtained for largest instances in each of the experiments.

\begin{table}
 \caption{Processing times for largest instances in experiments}\label{tab:processingTimes}
 \centering
 \begin{tabular}{|l|c|c|c|}
 \hline
   & \multicolumn{3}{|c|}{Implementation [ms]} \\
\cline{2-4}
  Algorithm (instance size) & GPU & SSE & Sequential \\
 \hline
 PLUGIN ($n=32768$) & 87.9 &  1442.3 & 47435.3\\
 LSCV\_h ($n=1024$, $d=16$) & 14.7 & 344.1 & 8283.6 \\
 LSCV\_H ($n=16384$, $d=16$) & 184.2 & 2320 & 53258.8 \\
 \hline
 \end{tabular}

\end{table}

\section{Conclusion}\label{sec:c-fw}
In the paper we have presented some methods of how to efficiently compute the so called bandwidth parameter used in computing kernel probability density functions (KPDF). The functions can be potentially used in various database and data exploration tasks. One possible application is the task known as approximate query processing. However, the serious drawback of the KPDF approach is that computations of the bandwidth parameter  -- a crucial parameter in KPDF -- are very time consuming. To solve this problem we have investigated several methods of optimizing these computations. We utilized two SIMD architectures: \emph{SSE CPU} architecture and \emph{NVIDIA GPU} architecture to accelerate computations needed to find the optimal value of the bandwidth parameter. We have tested our SSE and GPU implementations using three classical algorithms for finding the optimal bandwidth parameter: PLUGIN, LSCV\_h and LSCV\_H. Detailed mathematical formulas are presented in section \ref{sec:form-band-sel}. As for LSCV\_h algorithm we have proposed some slight but important changes in the basic mathematical formulas. The changes allow us to precompute some values which may be later reused many times. The details can be found in section \ref{sec:math-modif}. The fast SSE and CUDA implementations are also compared with a simple sequential one. All the necessary details on using SIMD architectures for fast computation of the bandwidth parameter are presented in section \ref{sec:simd} and the final notes on how to utilize the algorithms are presented in section \ref{sec:algorithms}. Our GPU implementations were about 300-500 times faster than their sequential counterparts and 12-23 times faster than their SSE counterparts. SSE implementations were about 20-30 times faster than sequential implementations. The above results confirm the great usability of modern processing units. All the codes developed have been made public available and can be downloaded from \url{http://www.cs.put.poznan.pl/wandrzejewski/_sources/hcode.zip}.



\bibliographystyle{spmpsci}
\bibliography{paper}

\appendix

\section{Derivation of equations \eqref{eq:colInTriangleMatrix} and \eqref{eq:rowInTriangleMatrix}}\label{sec:jqderiv}

In this appendix we provide detailed derivation of equations \eqref{eq:colInTriangleMatrix} and \eqref{eq:rowInTriangleMatrix} used for calculating thread block's position within an upper triangular matrix. We assume that the thread blocks are started within a one dimensional grid and are assigned unique non negative consecutive numbers $bx$ (see Figure \ref{fig:bx2lq}). Based on $bx$ we want to compute block's column (denoted $l$) and block's row (denoted $q$) within an upper triangular matrix. 

We shall start with deriving the number of blocks $n_b$ contained in triangular matrix. As each consecutive column has one more block than the previous one (i.e. column number $l$ contains $l+1$ blocks) the number of blocks up to a column number $l$ (i.e. $n=l+1$ columns) may be calculated as a sum of arithmetic progression:
\begin{eqnarray}
 a_1 & = & 1, \\
 a_n & = & l+1, \\
 n_b & = & n(a_1+a_n)/2 = (l+1)(l+2)/2.  \label{eq:appendixA_nb}
\end{eqnarray}

\noindent  Let us now assume that we have a given $bx$ number of a block and we know that this block is on the main diagonal of the triangular matrix. We will now find the column number $l$ for this block. Let us consider a submatrix of the triangular matrix such that the $bx$-th block is in the lower right corner of this submatrix. As $bx$ numbers are zero-based we know that there must be $n_b=bx+1$ blocks in this sumbatrix. Let us substitute this to the equation~\eqref{eq:appendixA_nb}:

\begin{eqnarray}
 (l+1)(l+2)/2 & = & bx+1, \\
 l^2+3l-2bx & = & 0. \label{eq:appendixA_quadraticeq}
\end{eqnarray}

\noindent  We solve the quadratic equation \eqref{eq:appendixA_quadraticeq}:

\begin{eqnarray}
 \Delta & = & \sqrt{8bx+9}, \\
 l_1 & = & (-\sqrt{8bx+9}-3)/2, \label{eq:appendixA_l1} \\
 l_2 & = & (\sqrt{8bx+9}-3)/2. \label{eq:appendixA_l2} 
\end{eqnarray}

\noindent As the result of equation \eqref{eq:appendixA_l1} is always negative the column number we are searching for is given by equation \eqref{eq:appendixA_l2}. 

Let us now assume that $bx$ block is not on the main diagonal. Let $bx_1$ be the biggest block number such that $bx_1<bx$ and let $bx_2$ be the smallest block number such that $bx<bx_2$. Let $l_{bx_1}$ be the column of $bx_1$ block  and $l_{bx_1}$ be the column of $bx_2$ block. It is easy to notic, that $l_{bx_1}+1=l_{bx_2}$. Consequently, the $l_2$ column calculated for $bx$ block using equation \eqref{eq:appendixA_l2} must be a non integer value $l_{bx_1}<l_2<l_{bx_2}$. Moreover, as $bx_1$ is the number of the last block in its column, $bx$ block must be in the same column as $bx_2$ block, i.e. its column number is the smallest integer value greater or equal to $l_2$ calculated using equation \eqref{eq:appendixA_l2}. This leads to the equation \eqref{eq:colInTriangleMatrix}:

$$
 l=\left\lceil\frac{\sqrt{8bx+9}-3}{2} \right\rceil.
$$

To find block's row number $q$, we need to subtract from $bx$ the number of blocks in previous columns. This can be easily found by substituting $l-1$ into equation \eqref{eq:appendixA_nb}, which leads to the final equation \eqref{eq:rowInTriangleMatrix}:

$$
 q=bx-\frac{l(l+1)}{2}.
$$

\end{document}